\documentclass{jfm}
\usepackage{graphicx}
\usepackage{epstopdf, epsfig}
\usepackage{color}
\usepackage{ulem}

\usepackage{amsmath}
\usepackage{hyperref}
\hypersetup{
    colorlinks = true,
    urlcolor   = blue,
    citecolor  = blue,
}
\shorttitle{Turbulence modulation in liquid-liquid two-phase TC flow}
\shortauthor{J Su, C Wang, Y Zhang, F Xu, J Wang, and C Sun}

\title{Turbulence modulation in liquid-liquid two-phase Taylor-Couette turbulence}

\author{Jinghong Su\aff{1},
  Cheng Wang\aff{1},
  Yi-Bao Zhang\aff{1},
  Fan Xu\aff{4},
  Junwu Wang\aff{2,4}
  \corresp{\email{jwwang@cup.edu.cn}},
  Chao Sun\aff{1,3}\corresp{\email{chaosun@tsinghua.edu.cn}}}

\affiliation{\aff{1}New Cornerstone Science Laboratory, Center for Combustion Energy, Key Laboratory for Thermal Science and Power Engineering of Ministry of Education, Department of Energy and Power Engineering, Tsinghua University, 100084 Beijing, China
\aff{2}Beijing Key Laboratory of Process Fluid Filtration and Separation, College of Mechanical and Transportation Engineering, China University of Petroleum, Beijing 102249, P R. China
\aff{3}Department of Engineering Mechanics, School of Aerospace Engineering, Tsinghua University, Beijing 100084, P. R. China
\aff{4}State Key Laboratory of Mesoscience and Engineering, Institute of Process Engineering, Chinese Academy of Sciences, P. O. Box 353, Beijing 100190, P. R. China}

\begin{document}

\maketitle

\begin{abstract}

We investigate the coupling effects of the two-phase interface, viscosity ratio, and density ratio of the dispersed phase to the continuous phase on the flow statistics in two-phase Taylor-Couette turbulence at a system Reynolds number of $6\times10^3$ \textcolor{black}{and a system Weber number of 10} using interface-resolved three-dimensional direct numerical simulations with the volume-of-fluid method. Our study focuses on four different scenarios: neutral droplets, low-viscosity droplets, light droplets, and low-viscosity light droplets.
We find that neutral droplets and low-viscosity droplets primarily contribute to drag enhancement through the two-phase interface, while light droplets reduce the system’s drag by explicitly reducing Reynolds stress due to the density dependence of Reynolds stress. Additionally, low-viscosity light droplets contribute to greater drag reduction by further reducing momentum transport near the inner cylinder and implicitly reducing Reynolds stress.
While interfacial tension enhances turbulent kinetic energy (TKE) transport, drag enhancement is not strongly correlated with TKE transport for both neutral droplets and low-viscosity droplets. \textcolor{black}{Light droplets primarily reduce the production term by diminishing Reynolds stress, whereas the density contrast between the phases boosts TKE transport near the inner wall. Therefore, the reduction in the dissipation rate is predominantly attributed to decreased turbulence production, causing drag reduction.}
For low-viscosity light droplets, the production term diminishes further, primarily due to their greater reduction in Reynolds stress, while reduced viscosity weakens the density difference's contribution to TKE transport near the inner cylinder, resulting in a more pronounced reduction in the dissipation rate and consequently stronger drag reduction.
Our findings provide new insights into the physics of turbulence modulation by the dispersed phase in two-phase turbulence systems.

\end{abstract}


\section{Introduction}
Two-phase liquid-liquid flows, often referred to as emulsions, involve the presence of two immiscible liquids and play a significant role in various natural and industrial processes. Turbulent emulsions, in particular, introduce further complexity due to the presence of dispersed phases. The interaction between the continuous phase and the dispersed phase leads to additional complexities, resulting in a wide range of phenomena and dynamics influenced by the properties of the phases and the explored parameter regimes~\citep{lemenand2017turbulent, yi2023recent}. In recent decades, turbulent emulsions have found extensive applications in fields such as petroleum, food, pharmaceuticals, and cosmetics~\citep{spernath2006microemulsions, wang2007oil, mcclements2007critical, kilpatrick2012water}, garnering significant interest~\citep{rosti2018rheology, yi2023recent, ni2023deformation}. However, our understanding of how turbulence is influenced by the dispersed phase in turbulent emulsions remains limited.

Due to the presence of the two-phase interface and disparities in liquid properties between the two phases, experimental observation of behavior in turbulent emulsions becomes particularly challenging. Current experimental research primarily focuses on microscopic droplet formation, size distribution, and the macroscopic response of global transport~\citep{bakhuis2021catastrophic, yi2021global, wang2022turbulence, yi2022physical}. Meanwhile, related simulations have been employed to investigate droplet size distribution, primarily in homogeneous and isotropic turbulence, with a specific emphasis on how turbulence affects droplet breakup behavior~\citep{mukherjee2019droplet, vela2022memoryless}.
The feedback of droplets on turbulence has recently been studied in homogeneous shear turbulence.\textcolor{black}{~\cite{dodd2016interaction} investigated how droplet deformation, breakup and coalescence affect in the temporal evolution of turbulent kinetic energy (TKE). They showed that droplet coalescence reduces the total interfacial surface area, causing a decrease in surface energy and an increase in local kinetic energy.} The presence of droplets acts as a sink in the TKE of the bulk fluid, as the dispersed phase was found to slow down the dissipation of TKE compared to the continuous phase~\citep{rosti2019droplets}.
The effect of droplets and the role of their viscosity on turbulence in homogeneous and isotropic turbulent flows have also received recent attention. The energy is transported consistently from large to small scales by the two-phase interface, and the total interface area is directly proportional to the amount of energy transported. Increasing the dispersed phase viscosity would reduce the amount of energy being transported~\citep{crialesi-esposito_rosti_chibbaro_brandt_2022}.
Correspondingly, large velocity gradients are found across the two-phase interface and will gradually disappear as the viscosity of the dispersed phase increases~\citep{farsoiya2023role}. These findings provide a deeper understanding of the impact of droplets on homogeneous turbulence. However, since the studied systems are unbounded, the results cannot be directly applied to wall-bounded turbulence, where strong inhomogeneity and anisotropy could develop due to the boundary layer. This introduces the potential for different observations in both phases.

In contrast to the relatively uniform turbulence dissipation observed throughout different regions in homogeneous turbulence, wall-bounded turbulence exhibits a distinct characteristic where approximately half of the dissipation occurs predominantly in the immediate vicinity of walls~\citep{jimenez2012cascades}.
Recent studies in wall-bounded laminar flows have investigated the modulation of the two-phase interface on the system’s drag \textcolor{black}{by allowing droplets to coalesce numerically} in the Taylor-Couette device~\citep{hori2023interfacial} or prohibiting droplet coalescence in planar Couette flow~\citep{de2019effect}. The drag enhancement caused by the dispersed phase is attributed to the interfacial contribution, and coalescence could effectively decrease the interfacial area, thus weakening the drag enhancement effect.
To investigate drag modulation by different dispersed phases, our previous work~\citep{su2024numerical} examined turbulence modulation induced by dispersed phases with varying density and viscosity through three-dimensional direct numerical simulations in turbulent Taylor-Couette flow. We derived a momentum transport formula, revealing that the two-phase interface consistently enhances drag. Reducing the density or viscosity of the dispersed phase decreases the contributions of the advection and diffusion terms, leading to reduced drag. However, that work only encompassed the effects of the dispersed phase on global properties, leaving an unexplored area in understanding the effect on the local statistical properties of turbulent flow. The impact of the dispersed phase on the production and dissipation of turbulence remains unknown. This motivates further exploration in this new work to gain a comprehensive understanding of the modulation of the transport of TKE by examining the effects of the dispersed phase on the local statistical properties.

In this work, we conducted a comprehensive study on the modulation of statistical properties of turbulence induced by dispersed phases in the Taylor-Couette system. To achieve this, we employed an interface-resolved volume-of-fluid method, which allows us to resolve the interface between the two phases and solve the governing equations in a single-equation formulation. This approach enables us to perform operations similar to those in single-phase flow, facilitating effective comparisons and determining the specific effects of the dispersed phase.

We aim to explore the behavior of turbulent emulsions in a semi-dilute regime, with a specific focus on the potential impact of droplets possessing lower density and viscosity compared to the continuous phase. Through numerical simulations, we can disentangle the effects of the two-phase interface, the density, and the viscosity of the dispersed phase on turbulence modulation, uncovering their intertwined coupling.
The manuscript is organized as follows: In \S 2, the numerical method and settings are described. In \S 3, turbulence modulation is discussed based on momentum budget analysis, turbulence fluctuation analysis, and TKE budget analysis. Finally, conclusions are drawn in \S 4.


\section{Numerical method and setting}\label{sec2}
We conducted interface-resolved three-dimensional direct numerical simulations to investigate the two-phase fluid-fluid turbulence in a Taylor-Couette (TC) system. These simulations were performed using a volume-of-fluid (VOF) method with a piecewise-linear interface calculation (PLIC) algorithm implemented in the interFoam solver of the open-source OpenFOAM v8~\citep{rusche2003computational,chen2022turbulent}. The robustness of OpenFOAM in simulating single-phase TC turbulence and two-phase TC turbulence has been demonstrated in our previous works~\citep{xu2022direct, xu2023direct,su2024numerical}.

We consider two immiscible and incompressible fluids confined between two coaxial cylinders whose radii are $r_i$ (inner) and $r_o$ (outer). In this work, we fix the outer cylinder while allowing the inner cylinder to rotate with a constant angular velocity $\omega_i$.
The two-phase incompressible flow is governed by the Navier-Stokes equations 
\begin{equation}
  \nabla\cdot\boldsymbol{u}
  =0,
  \vspace{-0mm}
\end{equation}
\begin{equation}
  {\partial _t(\rho \boldsymbol{u})}
  +\nabla\cdot(\rho\boldsymbol{uu})
  =
  -{\nabla{p}}
  +\nabla\cdot{\boldsymbol{\tau}}
  +\boldsymbol{f},
\end{equation}
where $\boldsymbol{u}$ is the velocity field, $p$ is the pressure and $\boldsymbol{\tau} =\mu(\nabla\boldsymbol{u}+(\nabla{\boldsymbol{u}})^T)$ is the viscous stress. $\rho$ and $\mu$ are the density and viscosity of the combined phase. The phase fraction $\alpha$ is introduced to characterize the variable density and viscosity, i.e., $\rho=\alpha\rho_d+(1-\alpha)\rho_f $ and $\mu=\alpha\mu_d+(1-\alpha)\mu_f $, where $\rho$ and $\mu$ with subscripts $f$ and $d$ stand for the density and viscosity of continuous phase and dispersed phase. 
The continuum surface force method, as proposed by \cite{brackbill1992continuum}, is adopted in this study to describe the interfacial tension, i.e., \textcolor{black}{$\boldsymbol{f}=\sigma\kappa\nabla\alpha$}, where $\sigma$ denotes the surface tension coefficient and $\kappa=-\nabla\cdot(\nabla\alpha/|\nabla\alpha|)$ represents the interface curvature.

In the VOF method, the phase fraction $\alpha$ is utilized in each cell to characterize the distribution of the two phases. The range of $\alpha$ is from zero to one, where $\alpha=0$ represents the continuous phase, $\alpha=1$ represents the dispersed phase, and $0<\alpha<1$ represents the interface region.
The evolution of $\alpha$ is governed by the transport equation
\begin{equation}
  {\partial _t \alpha}
  +\nabla\cdot(\alpha \boldsymbol{u})
  =0.
  \vspace{-0mm}
\end{equation}
Because of the continuity of the phase fraction, the interface between the two phases tends to become smeared. To mitigate this issue, a PLIC-based algorithm has been recently implemented to capture the interface accurately. This algorithm involves representing the interface between the two phases by employing surface-cuts, which split each cell to match the phase fraction in that cell. \textcolor{black}{The surface-cuts are oriented according to the phase fraction gradient.}
The phase fraction on each cell face is then calculated from the amount submerged below the surface cut. Based on this algorithm, the resolved interface region ($0<\alpha<1$) could be confined within a single layer of grid cells between the two phases to ensure the sharpness of the interface. It's important to note that this algorithm may encounter difficulty handling certain cells when the cut position is unclear or when multiple interfaces exist. In such cases, the interface compression approach proposed by \cite{weller2008new} is applied to those cells. In this approach, an artificial compression term, which is only active in the vicinity of the interface, is added to the transport equation to prevent interface smearing based on counter-gradient transport, i.e.,
\begin{equation} 
  {\partial _t \alpha}
  +\nabla\cdot(\alpha \boldsymbol{u})
  +\nabla\cdot[\alpha(1-\alpha) \boldsymbol{u}_c]
  =0,
  \vspace{-0mm}
\end{equation}
where $\boldsymbol{u}_c = c\boldsymbol{u}\nabla{\alpha}/|\nabla{\alpha}|$ with $c$ being the compression factor. In addition, the multidimensional universal limiter with explicit solution (MULES) algorithm is implemented to ensure that the phase fraction $\alpha$ remains within the strict bounds of 0 and 1. The combination of the PLIC-based algorithm with the interface compression approach allows the present approach to be easier to implement even with an unstructured mesh, thereby increasing the robustness of the solutions. Therefore, this PLIC-based VOF method has been applied in our study to deal with the two-phase turbulence in the TC system.

To minimize computational costs without compromising the accuracy of our results, we selected a rotational symmetry of order 6 (i.e., the azimuthal angle of the simulated domain is $\pi/3$) and an aspect ratio of $\Gamma=L/d=2\pi/3$ in the simulated Taylor-Couette system, where $d$ corresponds to the gap width between the cylinders and $L$ represents the axial length. This choice has been validated for both single-phase and multiphase Taylor-Couette turbulence~\citep{brauckmann2013direct,spandan2018physical,assen2022strong}.  The curvature of the Taylor-Couette system is characterized by the ratio $\eta=r_i/r_o=0.714$. 
The Taylor number, denoted as $Ta =  \chi(r_o+r_i)^2(r_o-r_i)^2\omega_i^2/(4\nu_f^2)$, is fixed as $5.49\times 10^7$, where $\chi = [(r_i+r_o)/(2\sqrt{r_i r_o})]^4$ and $\nu_f=\mu_f/\rho_f$. The system Weber number, denoted as $We$, is given by ${\rho_f}u_{\tau}^2 d/\sigma$, with a fixed value of 10. The $u_{\tau}$ is the friction velocity defined as $\sqrt{\tau_w/\rho_f}$, where $\tau_{w}$ represents the shear stress at the inner wall for single-phase flow. The frictional Reynolds number at the inner cylinder, denoted as $ Re_\tau={\rho_f} u_{\tau} d/\mu _f$, is fixed as 295.39. The system Reynolds number ${\rm Re}={\rho_f} u_i d/\mu_f$ is fixed as 6000, where $u_i = r_i \omega_i$ is the velocity of the inner cylinder.

Physically, the length scales of fluid-fluid two-phase turbulence involved can range from the largest scale of the problem down to the Kolmogorov scale of turbulence, and even further to the molecular scale of coalescence and breakup events at the interface. Specifically, during the coalescence event, the length scales associated with film drainage can be as small as a few hundred nanometers or less~\citep{kamp2017drop}. Ideally, it would be advantageous to conduct simulations that fully resolve all scales, similar to the study of single-phase turbulence. However, this approach is not feasible for fluid-fluid two-phase turbulence due to the significant separation between the largest flow scale and the smallest interfacial scale, which can span up to eight to nine orders of magnitude. Such a wide range of scales would require tremendous computational resources. As a result, the conventional approach is to avoid resolving the molecular scales at the interface and instead focus on resolving all turbulence scales, from the larger macroscopic scale down to the Kolmogorov length scale~\citep{soligo2019breakage}. 
\textcolor{black}{In the VOF method, the coalescence and breakup of droplets is handled implicitly and two separate interfaces automatically merge when they occupy the same computational cell. This process is commonly referred to as numerical coalescence. Given that fully resolving film drainage and turbulence is prohibitively expensive from a computational point of view. An alternative approach is to use a subgrid-scale model to determine whether the droplets will coalesce. However, such approaches are highly dependent on the underlying film drainage model used and therefore their predictive capabilities are uncertain. As noted in the review of~\cite{soligo2021turbulent}, there has been no fully validated method to accurately model film drainage in two-phase turbulence. In our simulations we do not use a film drainage model and allow the droplets to numerically coalesce, which is generally acceptable in two-phase flow simulations in dilute and semi-dilute regimes~\citep{rosti2019droplets,crialesi-esposito_rosti_chibbaro_brandt_2022}.}

No-slip and impermeable boundary conditions are imposed in the radial direction, while periodicity is imposed in the axial and azimuthal directions. The inner and outer cylinders are subjected to a Neumann boundary condition for the phase fraction, resulting in a default contact angle of $90^\circ$. The maximum Courant-Friedrichs-Lewy number is set to be 0.2. 
Firstly, a single-phase case is simulated to initialize the velocity field. Once a well-developed flow with a pair of Taylor rolls is obtained, the simulation is restarted, in which the spheres of diameter $0.2d$ containing the dispersed phase are uniformly positioned in the domain. Two different volume fractions of the dispersed phase, $\varphi = 5\%$ and $\varphi = 10\%$, are considered. The dispersed phase undergoes continuous coalescence and breakup, gradually adapting to the flow field. 
\textcolor{black}{All the presented statistics are collected for at least $3 \times 10^2$ large eddy turnover times ($(r_o-r_i)/(\omega_i r_i)$) after the two-phase flow reaches a statistically steady state. }

The Taylor-Couette system is discretized using a collocated grid system consisting of $N_\theta \times N_r \times N_z = 336\times250\times192$ grids in the azimuthal, radial, and axial directions, respectively. The grids are distributed uniformly in the azimuthal and axial directions but are unevenly spaced and concentrated near the two cylinders in the wall-normal direction. The grid spacing is measured in units of the viscous length scale $\delta_\nu = \nu_f/u_\tau$ for single-phase flow. In the radial direction, the grid spacing varies from $0.34\delta_\nu$ near the wall to $2.73\delta_\nu$ at the center of the gap. In the azimuthal direction, it ranges from $2.3\delta_\nu$ near the inner wall to $3.22\delta_\nu$ near the outer cylinder. The grid spacing remains uniform in the axial direction, with a value of $3.22\delta_\nu$.
The Kolmogorov scale for single-phase flow, denoted as $\eta_k$, is determined to be $2.11\delta_\nu$ by employing the exact dissipation relationships given by $\eta_k/d=({\chi}^{-2}Ta(Nu_{\omega}-1))^{-1/4}$, where $Nu_\omega=T/T_{lam}$~\citep{ eckhardt2007torque} with $T$ representing the torque required to drive the cylinders and $T_{lam}$ corresponding to the torque when the flow is purely laminar. The maximum grid spacing is constrained to be nearly $1.5\eta_k$ to ensure that the grid length remains comparable to the scale of the local Kolmogorov length.

\textcolor{black}{We utilize a blended scheme with the blending factor being 0.9 for the temporal term discretization, which lies between the first-order Euler scheme and the second-order Crank-Nicolson scheme. The introduction of the dispersed phase could lead to numerical instability. The blended scheme ensures a balance between numerical stability and numerical accuracy~\citep{greenshields2020}. For spatial discretization, we employ a second-order linear-upwind scheme to discretize the advection term in the momentum equation.
The PIMPLE algorithm~\citep{holzmann2016mathematics}, which is a hybrid version of the PISO algorithm and the SIMPLE algorithm, is used to handle the pressure–velocity coupling. The PIMPLE algorithm guarantees better stability for problems that involve very large timesteps and pseudo-transient simulation. The pressure equation is solved using the Geometric Algebraic Multigrid (GAMG) solver coupled with the Simplified Diagonal-based Incomplete Cholesky (DIC). The GAMG leverages the multigrid approach, which utilizes a hierarchy of grids with different resolutions to accelerate the convergence process. This allows GAMG to quickly converge to a solution, reducing computational time compared to other standard methods. In OpenFOAM, the GAMG is commonly coupled with DIC to speed up the computational efficiency in simulating two-phase flow~\citep{scheufler2019accurate,chen2022turbulent}. 
For solving velocity and phase fraction, we use an iterative solver with a symmetric Gauss-Seidel smoother. The Gauss-Seidel method is known for several advantages over other techniques. Its convergence speed and memory efficiency are particularly noteworthy.
In the simulation, we maintain a tolerance of $10^{-6}$ for all variables to control the residuals, except for the phase fraction, which has a tolerance of $10^{-8}$. The computational accuracy of these settings is examined and validated in appendix~\ref{appA} and~\ref{appB}, demonstrating that our methods can effectively simulate single-phase and two-phase cases.}

\section{Results}
To examine the impact of various parameters on drag modulation, we conducted a comprehensive investigation by sequentially altering the volume fraction $\varphi$, viscosity ratio $\xi_{\mu}$, and density ratio $\xi_{\rho}$ to study their individual and coupled effects, as outlined in table~\ref{tab:drag}.
\textcolor{black}{In this work, we fix the outer cylinder while sustaining the constant rotational velocity of the inner cylinder. The torque $T$ required to drive the inner cylinder is examined to study the drag modulation caused by different types of droplets. 
This setup is commonly used to study drag modulation caused by droplets, bubbles, and particles~\citep{spandan2016drag,bakhuis2018finite,yi2021global,yi2022physical}.}
Our work focuses on four typical cases, including two-phase flows with the dispersed phase being neutral droplets ($\xi_{\rho} = 1$ and $\xi_{\mu} = 1$), low-viscosity droplets ($\xi_{\rho} = 1$ and $\xi_{\mu} = 1/4$), light droplets ($\xi_{\rho} = 1/4$ and $\xi_{\mu} = 1$), and low-viscosity light droplets ($\xi_{\rho} = 1/4$ and $\xi_{\mu} = 1/4$). \textcolor{black}{The torque is collected from about $10^5$ time steps after reaching a statistically steady state.}
An increase in drag is observed for two-phase flow with neutral droplets, consistent with the experimental observation~\citep{yi2021global}. However, reducing the viscosity of the dispersed phase to $\xi_{\mu} = 1/4$ results in almost no change in the drag enhancement effect compared to the case with neutral droplets. On the other hand, reducing the density of the dispersed phase to $\xi_{\rho} = 1/4$ leads to significant drag reduction. Moreover, simultaneously reducing the density and the viscosity can result in stronger drag reduction. \textcolor{black}{For low-viscosity light droplets, the kinematic viscosity is the same as that of the continuous phase. Therefore, the corresponding characteristic Reynolds number $Re_d = {\rho_d} u_i d/\mu_d=6000$ is the same as the system Reynolds number for the single-phase case. Considering the same Reynolds number and a viscosity ratio of 1/4, a drag reduction of $75\%$ would be obtained when the total volume fraction of the low-viscosity light droplets is $100\%$. In our work, only $10\%$ low-viscosity light droplets could lead to up to $30\%$ drag reduction, demonstrating the efficiency of the chosen volume fraction on drag modulation.}
Additionally, we observe similar trends in drag modulation for two different volume fractions of the dispersed phase, $\varphi = 5\%$ and $\varphi = 10\%$. For given density ratio $\xi_{\rho}$ and viscosity ratio $\xi_{\mu}$, an increase in the volume fraction $\varphi$ leads to the pronounced amplitude of drag enhancement or reduction, which depends on the droplet properties ($[\xi_{\rho}$, $\xi_{\mu}]$). \textcolor{black}{We will conduct a detailed analysis of these results to elucidate the individual and coupled effects of the viscosity and density of the dispersed phase on the system's drag and turbulence properties, aiming to reveal the underlying mechanisms.}

Given the similarity in drag modulation for the two volume fractions, $\varphi = 5\%$ and $\varphi = 10\%$, we will focus our subsequent investigations on the $5\%$ volume fraction cases.
In a TC system, the spatial distribution of droplets is affected by the background flow field and potentially by the inertial effect due to centrifugal force if there is a density mismatch between the phases. Figures~\ref{fig:snapshots}a--\ref{fig:snapshots}h display the instantaneous interface snapshots as well as the azimuthally- and time-averaged phase fraction. The averaged radial-axial velocity vectors denote the magnitude and structure of the Taylor-vortex, which provides information on the dispersed phase spatial distribution. 

\subsection{Drag modulation}
\begin{table}
    	\centering
     \setlength\tabcolsep{16pt}
		\begin{tabular}{cccc}
			\textrm{$\varphi$}&
			\textrm{$\xi_{\rho}=\rho_d /\rho_f$}&
			\textrm{$\xi_{\mu}= \mu_d /\mu_f$}&
			\multicolumn{1}{c}{drag modulation ($T/T_{\varphi=0}-1$)}\\[6pt]
			$0  $  & ---		  & ---	 		 & ---		              \\[2pt]
			$5\%$ & $1$	  & $1$	 	& $+ 7.06\%$ 	      \\[2pt]
   	  $5\%$ & $1$	  & $1/4$	  & $+6.31\%$ 	      \\[2pt]
			$5\%$ & $1/4$ & $1$	 	& $-4.97\%$	       \\[2pt]
			$5\%$ & $1/4$ & $1/4$	 	& $-13.29\%$ 	       \\[2pt]
			$10\%$ & $1$	  & $1$	 	& $+9.79\%$  	       \\[2pt]
      	$10\%$ & $1$	  & $1/4$	   & $+8.56\%$ 	      \\[2pt]
			$10\%$ & $1/4$ & $1$	 	& $-16.28\%$ 	       \\[2pt]
			$10\%$ & $1/4$ & $1/4$	 	& $-29.76\%$ 	       \\[2pt]
		\end{tabular}
  \caption{\label{tab:drag} Drag modulation in two-phase flow. The $T$ represents the torque exerted on the inner cylinder and $T_{\varphi=0}$ denotes the torque specifically associated with the single-phase flow condition.}
\end{table}

\begin{figure*}
\centerline{\includegraphics[width=1\textwidth]{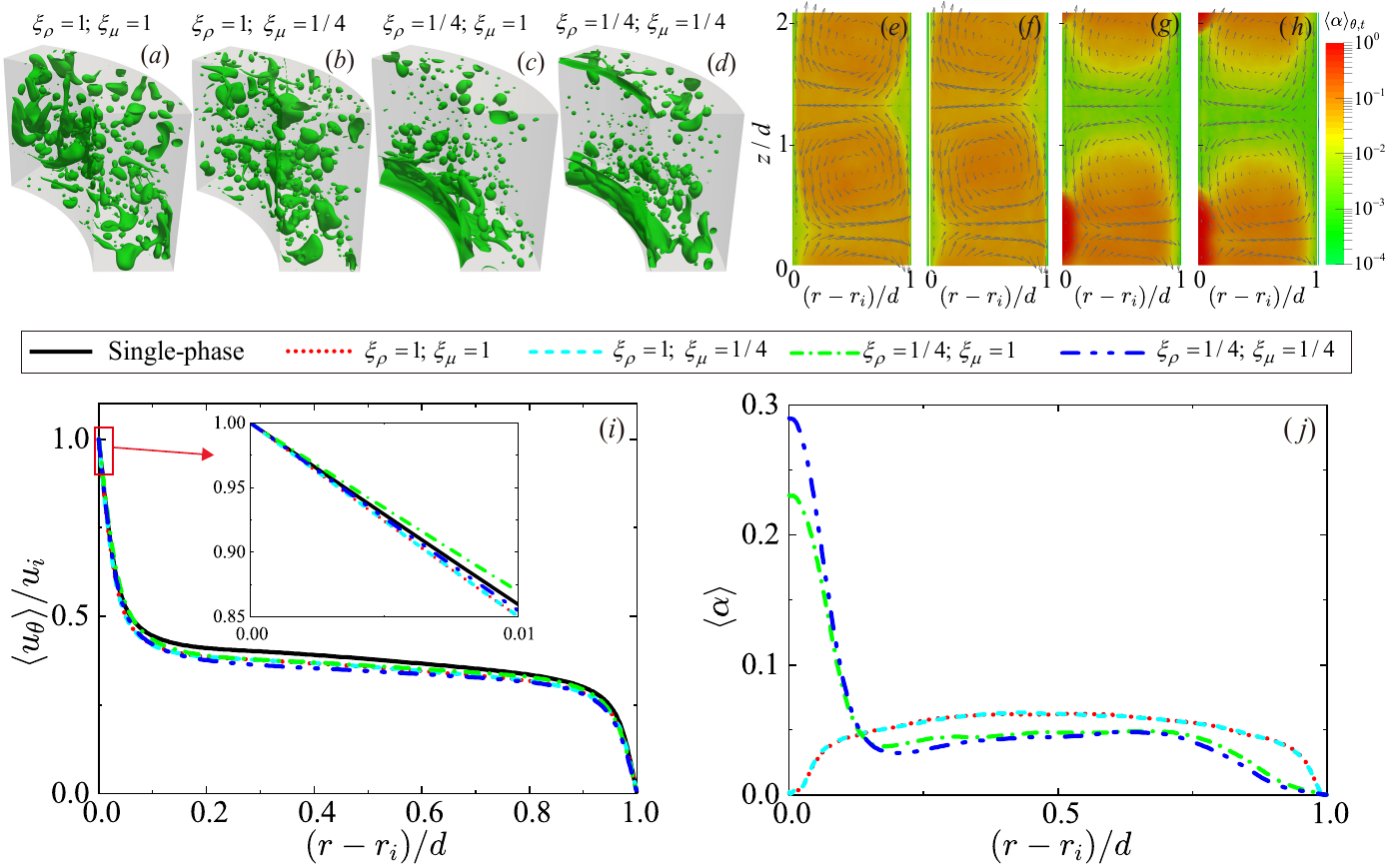}}%
  \caption{(\textit{a}-\textit{d}) Instantaneous interface snapshots and (\textit{e}-\textit{h}) corresponding azimuthally- and time-averaged phase fraction $\left \langle \alpha \right \rangle _{\theta,t}$. The grey arrows in (\textit{e}-\textit{h}) denote the direction and magnitude of the averaged radial-axial velocity vectors $\left \langle u_{rz} \right \rangle _{\theta,t}/u_i$. The $\left \langle \alpha \right \rangle _{\theta,t}$ indicates the proportion of dispersed phase (phase distribution). To facilitate the observation of the phase distribution, a logarithmic scale is employed for the legend labels. (\textit{i}) The normalized azimuthal velocity $\left \langle u_{\theta} \right \rangle/u_i$ and (\textit{j}) the phase distribution $\left \langle \alpha \right \rangle$ are obtained as a function of radial position $r$. Inset: zoom-in of the $\left \langle u_{\theta} \right \rangle/u_i$ near the inner cylinder. The operator $\left \langle \cdot \right \rangle$ denotes the average in the axial and azimuthal directions and over time.
 }
\label{fig:snapshots}
\end{figure*}

For neutral droplets, the effect of centrifugal force is eliminated as the densities of the two phases are perfectly matched. As illustrated in figure~\ref{fig:snapshots}(e), the neutral droplets predominantly exhibit voids in the region where the plumes are ejected from the inner (outer) boundary layer to the bulk. This pattern resembles that of neutrally buoyant finite-size particles at a system Reynolds number of 6500 in the same definition as here, which is primarily attributed to the flow structures and the finite-size effect of the particles~\citep{wang2022finite}. Although the neutral droplets have a weaker finite-size effect due to their capacity for deformation and breakage, the spatial distribution qualitatively aligns with that of finite-size particles. \textcolor{black}{Besides, compared to the single-phase case, the neutral droplets induce a decrease in the azimuthal velocity in the vicinity of the inner cylinder, as illustrated in the inset of figure~\ref{fig:snapshots}(i). Considering the fixed azimuthal velocity of the inner cylinder, this decrease causes an increase in the azimuthal velocity gradient at the inner cylinder, resulting in drag enhancement.} 

Low-viscosity droplets exhibit a distribution similar to that of the neutral droplets. This suggests that decreasing the viscosity of the dispersed phase does not introduce new mechanisms affecting the phase distribution, at least in the studied parameter regime. We note that the low-viscosity droplets are mainly distributed in the bulk region where the effect of viscosity on momentum transport is weak, which leads to a negligible effect of viscosity reduction on the flow field. \textcolor{black}{Therefore, when the droplets are neutrally buoyant, the decrease in viscosity ratio does not lead to a signiﬁcant change in the system's drag.}

For light droplets, the dispersed phase experiences two main forces within the $r-z$ plane. Firstly, the centrifugal force generated by the density difference between the two phases causes the dispersed phase to migrate toward the inner cylinder. \textcolor{black}{Secondly, the drag force, resulting from any velocity difference between the dispersed phase and the continuous phase, causes the dispersed phase to move along with the motion of the Taylor vortex. Ultimately, the dispersed phase tends to aggregate in the region close to the inner cylinder, where the drag and centrifugal forces are roughly balanced.} Remarkably, a prominent interfacial structure forms in the ejection region of the plumes near the inner cylinder, and the average phase fraction $\left \langle \alpha \right \rangle _{\theta, t}$ in this region is approximately 1 (see figures~\ref{fig:snapshots}c and~\ref{fig:snapshots}g). Consequently, there is a notable disparity in the distribution of the phase fraction near the inner cylinder compared to other regions. \textcolor{black}{In addition, an increase in the azimuthal velocity compared to the single-phase case is observed in the vicinity of the inner cylinder. Considering the fixed azimuthal velocity of the inner cylinder, this leads to a decrease in the azimuthal velocity gradient at the inner cylinder, which may be due to the density reduction lowering the effective Reynolds number. The decreased azimuthal velocity gradient at the inner cylinder results in drag reduction.}

For low-viscosity light droplets, the alteration from $\xi_{\mu} = 1$ to $\xi_{\mu} = 1/4$ disrupts the balance between the drag force and the centrifugal force, leading to a modification in the phase distribution. As depicted in figure~\ref{fig:snapshots}(h), the interfacial structure undergoes a slight stretching in the vertical direction, indicating a greater concentration of the dispersed phase near the inner cylinder, as depicted in figure~\ref{fig:snapshots}(j). This change can be attributed to the reduced viscosity, which weakens the Taylor-vortex and in turn makes the centrifugal force more significant. \textcolor{black}{It is important to note that, unlike light droplets, low-viscosity light droplets reduce the azimuthal velocity in the vicinity of the inner cylinder compared to the single-phase case, which results in a greater azimuthal velocity gradient. Therefore, the drag reduction induced by low-viscosity light droplets is mainly attributed to their low viscosity and little related to their modulation on the azimuthal velocity.}

\textcolor{black}{Based on the results discussed above, we found that neutral droplets mainly distribute in the bulk region. The presence of neutral droplets causes a decrease in the azimuthal velocity near the inner cylinder. This leads to an increase in the azimuthal velocity gradient at the inner cylinder, increasing the system's drag. Low-viscosity droplets exhibit similar distribution behavior to neutral droplets. As they are mainly distributed in the bulk region where the influence of viscosity is weak, the effect of viscosity reduction on azimuthal velocity is almost negligible. Therefore, low-viscosity droplets exhibit an indistinguishable decrease in drag compared to neutral droplets. 
In the case of light droplets, the centrifugal forces caused by density mismatch cause the light droplets to aggregate near the inner cylinder. The presence of light droplets leads to a decrease in the azimuthal velocity gradient at the inner cylinder probably due to that the light droplets lower the local effective Reynolds number, leading to drag reduction. 
For low-viscosity light droplets, a greater azimuthal velocity gradient is induced at the inner cylinder, which is contrary to the drag reduction caused by them. The drag reduction is mainly attributed to the low viscosity of the droplets which causes a reduction in the viscous shear stress at the inner cylinder. Additionally, it is observed in figure~\ref{fig:snapshots}(i) that the azimuthal velocity shows a reduction in the bulk region for all two-phase cases compared to the single-phase case. This phenomenon will be explained in the next section.}

\subsection{Momentum budget analysis}
To further investigate the turbulence modulation caused by different types of droplets, a momentum budget analysis is conducted based on the conserved quantity $J^\omega$ that characterizes the radial transport of azimuthal momentum in the two-phase Taylor-Couette turbulence~\citep{su2024numerical}
 \begin{equation}
  J^\omega=J_{adv}^\omega(r) + J_{dif}^\omega(r)+J_{int}^\omega(r),
  \label{equ:momentum transport}
\end{equation}
\textcolor{black}{where the three terms on the right-hand side represent the advection contribution, the diffusion contribution, and the interfacial contribution, respectively,} 
 \begin{equation}
 J_{adv}^\omega(r)=\left \langle r^2 \rho u_r u_\theta \right \rangle, 
   \label{equ:adv}
 \vspace{0mm}
 \end{equation}
\begin{equation}
J_{dif}^\omega(r)=-\left \langle \mu (r^3 \partial _r \omega + r\partial _\theta u_r) \right \rangle, 
\vspace{0mm}
 \end{equation}
\begin{equation}
 J_{int}^\omega(r)= -\int^r_{r_i} \left \langle {r^2 f_\theta}\right \rangle  \, dr.\\
\end{equation}
$u_r$ is the radial velocity, $\omega$ is the angular velocity and $f_\theta$ is the azimuthal component of interfacial tension. It is well-established that the conserved quantity $J^\omega$ and the torque at the inner cylinder $T$ in the Taylor-Couette system are related through the equation $T=2\pi L J^\omega$. \textcolor{black}{The advection contribution, the diffusion contribution, and the interfacial contribution are explicitly related to density, viscosity, and interfacial tension, respectively, offering a convenient way to effectively decouple the effects of density, viscosity, and two-phase interface on drag modulation.} 

\begin{figure*}
  \centerline{\includegraphics[width=1.00\textwidth]{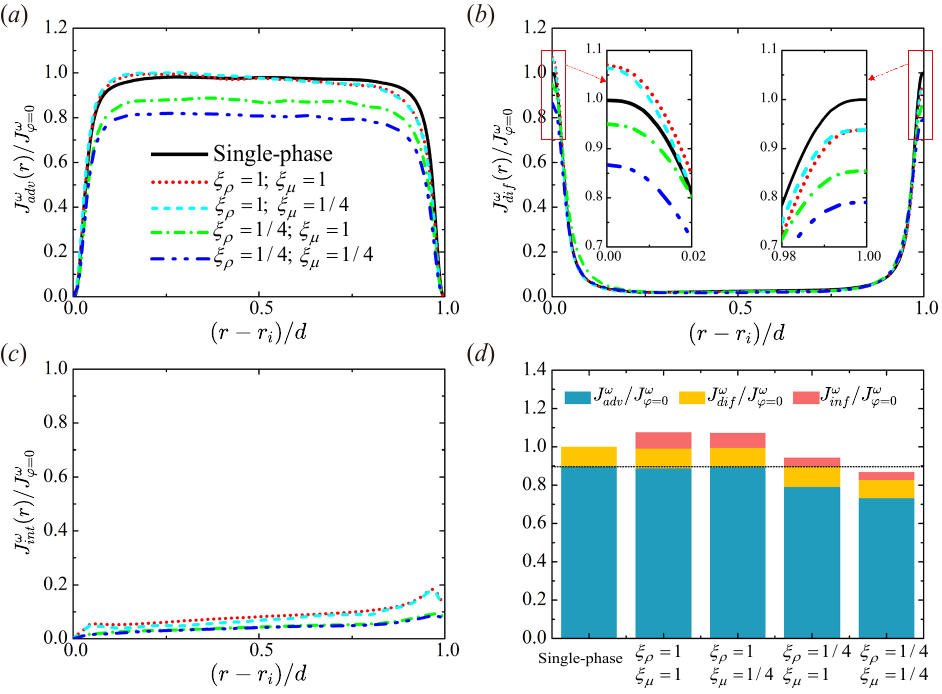}}%
  \caption{Momentum budget analysis. (\textit{a}) The advection contribution, (\textit{b}) the diffusion contribution, and (\textit{c}) the interfacial contribution as a function of the radial position are shown. (\textit{d}) The three contributions are averaged in the radial direction to characterize the corresponding terms within the whole system and are shown in the stack column. The dashed line in (\textit{d}) represents the averaged advection contribution for single-phase flow. All the quantities are normalized by the conserved quantity for single-phase flow. The insets in (\textit{b}) show the zoom-in view of the diffusion contribution near the inner and outer cylinders.
}
\label{fig:momentum transport}
\end{figure*}

For the two-phase flow with neutral droplets ($\xi_{\rho} = 1$ and $\xi_{\mu} = 1$), the advection contribution in the bulk region is modulated to account for the influence of the two-phase interface. Specifically, the advection contribution increases in the half of the bulk region closer to the inner cylinder, while decreasing in the other half of the bulk region (see figure~\ref{fig:momentum transport}a). Additionally, the presence of the two-phase interface introduces the interfacial contribution, which consistently exhibits a positive value along the radial position, indicating its contribution to drag enhancement (see figure~\ref{fig:momentum transport}c). Due to the lack of significant modulation in the radial-averaged advection and diffusion contributions, the increase in drag is primarily attributed to the interfacial contribution (see figure~\ref{fig:momentum transport}d). 

In contrast to other scenarios involving drag modulation (e.g., the drag enhancement caused by finite-size particles in turbulent channel flow and turbulent plane Couette flow, where the particle induced stress has zero contributions at the wall~\citep{picano2015turbulent,wang2017modulation}), the interfacial contribution in our cases exhibits a zero value at the inner cylinder and a non-zero value at the outer cylinder (\textcolor{black}{see figure~\ref{fig:momentum transport}c}). \textcolor{black}{Note that the interfacial contribution is obtained by integrating the contribution from the azimuthal component of interfacial tension within each cylindrical plane from the inner cylinder to the specific position. Namely, the interfacial contribution at a specific radial position relies on the interfacial tension in the region between the cylindrical plane and the inner cylinder, not just the interfacial tension within the cylindrical plane.}  

\textcolor{black}{It is found that the interfacial contribution shows an obvious increase with the radial position near the inner cylinder, while an obvious decrease with the radial position near the outer cylinder for the two-phase flow with neutral droplets.
Based on the distribution characteristics of the interfacial contribution and $J_{int}^\omega(r)= -\int^r_{r_i} \left \langle {r^2 f_\theta}\right \rangle  \, dr$, we propose a model to explain the effect of interfacial tension in shear turbulence. In this model, when the azimuthal velocity of the droplet phase is less than that of the surrounding continuous phase within a cylindrical plane, the droplet phase is subject to a drag force exerted by the surrounding continuous phase. The drag would be balanced by the interfacial tension 
$\left \langle f_\theta \right \rangle$, which takes a negative value and opposes the direction of the flow. Conversely, when the azimuthal velocity of the droplet phase is larger than that of the surrounding continuous phase, $\left \langle f_\theta \right \rangle$ takes a positive value and shares the same direction of the flow. In the vicinity of the inner cylinder, $\left \langle f_\theta \right \rangle$ takes a negative value and results in an obvious increase of interfacial contribution with the radial position. The interfacial tension opposes the direction of the flow and tends to reduce the azimuthal velocity as shown in the inset of figure~\ref{fig:snapshots}(i), thus leading to drag enhancement. Near the outer cylinder, $\left \langle f_\theta \right \rangle$ takes a positive value and results in an obvious decrease of interfacial contribution with the radial position, i.e., the interfacial tension tends to increase the azimuthal velocity there. The interfacial contribution can be regarded as the overall effect of interfacial tension in the region between a specific cylindrical plane and the inner cylinder. The positive interfacial contribution at the outer cylinder indicates that the overall effect of interfacial tension within the gap tends to reduce the azimuthal velocity. As there is a positive interfacial contribution at the outer cylinder for all two-phase cases, azimuthal velocity shows a reduction in the bulk region for all two-phase cases compared
to the single-phase case as shown in figure~\ref{fig:snapshots}(i).}

For the two-phase flow with low-viscosity droplets ($\xi_{\rho} = 1$ and $\xi_{\mu} = 1/4$), the three contributions are very close to those of neutral droplets since droplets mainly exist in the bulk region where the viscosity has a very weak influence on the momentum transport. Upon closer examination, it is still possible to glean some insights into the effects of viscosity. In comparison to neutral droplets, a minor increase in the radial-averaged advection contribution is observed, as depicted in figure~\ref{fig:momentum transport}(d). This could be attributed to the reduction in viscosity within the bulk region, leading to a diminished inhibitory effect of viscosity on turbulence within this region. Additionally, we observed that the radial-averaged diffusion contribution does not decrease despite the reduction in viscosity in the bulk region. This is because the decrease in droplet viscosity leads to an enhanced velocity gradient across the two-phase interface~\citep{farsoiya2023role}, which potentially contributes to the diffusion term (through $\partial _r \omega$ and $\partial _\theta u_r$). Moreover, reducing viscosity also decreases the interfacial contribution in the bulk region. We speculate that this is due to the viscosity reduction weakening the interface's ability to impede the surrounding flow field. 

For the two-phase flow with light droplets ($\xi_{\rho} = 1/4$ and $\xi_{\mu} = 1$), the advection contribution undergoes a significant reduction. The density ratio $\xi_{\rho} = 1/4$ promotes the aggregation of the dispersed phase near the inner cylinder, resulting in a substantial decrease in the upstream advection contribution due to its density-related nature. Moreover, in regions away from the inner cylinder where the dispersed phase's volume fraction is low, the significant reduction in the advection contribution can be attributed to a decrease in $u_r u_{\theta}$ since the density remains relatively constant. We here emphasize that the decrease in the advection contribution caused by $\xi_{\rho} = 1/4$ surpasses the interfacial contribution, leading to drag reduction. 

For the two-phase flow with low-viscosity light droplets ($\xi_{\rho} = 1/4$ and $\xi_{\mu} = 1/4$), the diffusion contribution is diminished near the inner cylinder due to its viscosity-related nature. \textcolor{black}{As discussed earlier regarding the azimuthal velocity, we have shown that the drag reduction is mainly due to the low viscosity of the droplets.} The advection contribution will be implicitly reduced through the velocity field (i.e., $u_ru_\theta$) to maintain the conserved quantity constant along the radial position.

\textcolor{black}{These results demonstrate that decreased viscosity alone does not significantly affect momentum transport, whereas the decreased viscosity in combination with decreased density can significantly reduce momentum transport and hence lead to drag reduction.}

\subsection{Turbulent fluctuation analysis}
The momentum budget analysis mainly focuses on the effects of interfacial tension and the fluid properties of the dispersed phase on global transport. However, while it sheds light on these aspects, the details of how the dispersed phase influences turbulence remain elusive. Therefore, it becomes crucial to delve deeper into statistics of turbulence properties.
\textcolor{black}{Due to the density difference between the dispersed and continuous phases, the Reynolds average is no longer the best choice for obtaining the Reynolds stress. Consequently, we adopt the Favre average method~\citep{favre1969statistical}, which applies a density-weighted average to the velocity. The Favre-averaged Navier-Stokes equations are formally similar to the Reynolds-averaged equations for the single-phase case and are easily compared across cases. This advantage makes the Favre average widely used in the study of compressible and multiphase flows.} The Favre average applies a density-weighted average to the velocity field, resulting in fluctuations in the velocity vector, 
 \begin{equation}
\boldsymbol{u}^{\prime\prime}
=
\boldsymbol{u}
-
\widetilde{\boldsymbol{u}}. 
  \label{equ:Favre}
\end{equation}
where 
$ \widetilde{\boldsymbol{u}}
=
\left \langle {\rho \boldsymbol{u}} \right \rangle
/
\overline{\rho}$ is the Favre-averaged velocity vector and $\overline{\rho}=\left \langle {\rho} \right \rangle$. 
On the other hand, the fluctuation in the velocity vector using the Reynolds average is given as 
 \begin{equation}
\boldsymbol{u}^{\prime}
=
\boldsymbol{u}
-
\left \langle {\boldsymbol{u}} \right \rangle.
\end{equation}
The velocity fluctuation due to the Favre average and the Reynolds average are related with
 \begin{equation}
\boldsymbol{u}^{\prime\prime}
=
\boldsymbol{u}^{\prime}
-
\boldsymbol{a},
  \label{equ:massflux}
\end{equation}
\textcolor{black}{where $\boldsymbol{a} = \left\langle {\rho}^{\prime}\boldsymbol{u}^{\prime} \right\rangle
/\left\langle {\rho} \right\rangle$ is a function of radial position, characterizing the overall effect of density fluctuations within each cylindrical plane.}
When the dispersed and continuous phases possess identical densities, the Favre average simplifies to the Reynolds average.

\begin{figure*}
  \centerline{\includegraphics[width=0.95\textwidth]{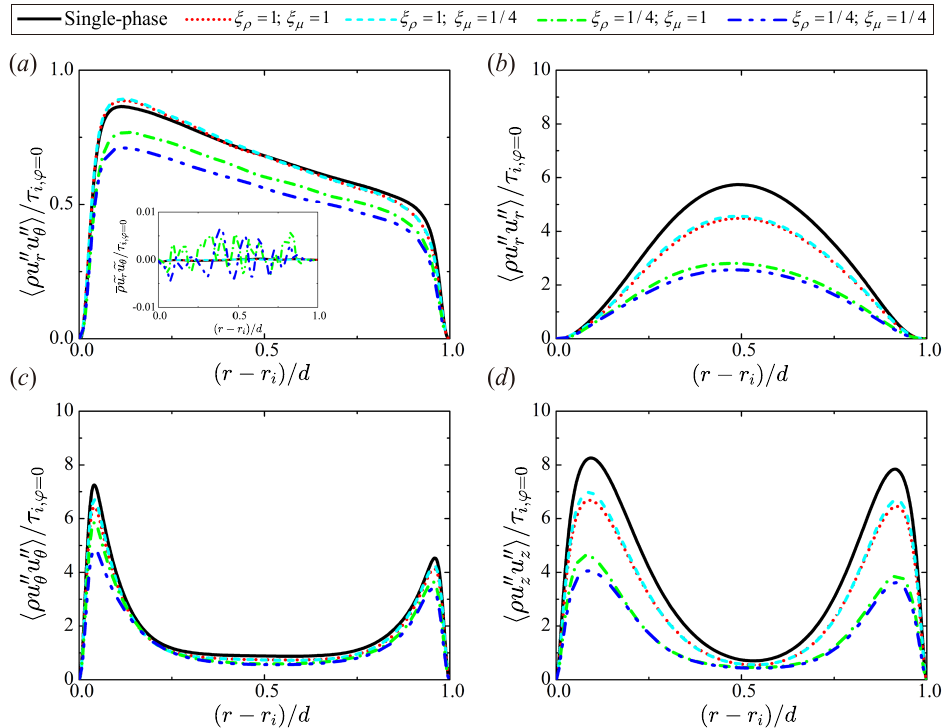}}%
  \caption{The modulation on the turbulence properties. 
  (\textit{a}) The Favre-averaged Reynolds stress 
  ${\left \langle {\rho u_r^{\prime\prime} u_\theta^{\prime\prime}} \right \rangle}/{\tau}_{i,\varphi=0}$, 
  (\textit{b}) the radial (wall-normal) component of the TKE 
   ${\left \langle {\rho u_r^{\prime\prime} u_r^{\prime\prime}} \right \rangle}/{\tau}_{i,\varphi=0}$, 
  (\textit{c}) the azimuthal (streamwise) component of the TKE 
   ${\left \langle {\rho u_\theta^{\prime\prime} u_\theta^{\prime\prime}} \right \rangle}/{\tau}_{i,\varphi=0}$, 
   (\textit{d}) and the axial (spanwise) component of the TKE 
  ${\left \langle {\rho u_z^{\prime\prime} u_z^{\prime\prime}} \right \rangle}/{\tau}_{i,\varphi=0}$ 
  are shown as a function of the radial position, where ${\tau}_{i,\varphi=0}$ is the total shear stress on the inner cylinder in single-phase flow. The inset figure in (\textit{a}) denotes the Favre-averaged term of the mean flow $\overline{\rho} \widetilde{u_r} \widetilde{u_\theta}/{\tau}_{i,\varphi=0}$. The maximum value of the ordinate of inset figure is set as 0.01, while the maximum value of the ordinate of (\textit{a}) is set as 1.}
\label{fig:fluctuation}
\end{figure*}

To explore the effect of the dispersed phase on the background turbulence, we show the Favre-average Reynolds stress 
${\left \langle {\rho u_r^{\prime\prime} u_\theta^{\prime\prime}} \right \rangle}$ 
as well as the radial component  
${\left \langle {\rho u_r^{\prime\prime} u_r^{\prime\prime}} \right \rangle}$, 
the azimuthal component  
${\left \langle {\rho u_\theta^{\prime\prime} u_\theta^{\prime\prime}} \right \rangle}$ 
and the axial component ${\left \langle {\rho u_z^{\prime\prime} u_z^{\prime\prime}} \right \rangle}$ of the TKE ${k=0.5\left \langle {\rho u_r^{\prime\prime} u_r^{\prime\prime}} + {\rho u_\theta^{\prime\prime} u_\theta^{\prime\prime}}+{\rho u_z^{\prime\prime} u_z^{\prime\prime}} \right \rangle}$ as depicted in figure~\ref{fig:fluctuation}. The Favre-average Reynolds stress (hereafter referred to as Reynolds stress)${\left \langle {\rho u_r^{\prime\prime} u_\theta^{\prime\prime}} \right \rangle}$  shows a similar profile to the advection contribution $\left \langle r^2 \rho u_r u_\theta \right \rangle$ in the momentum transport due to 
 \begin{equation}
\left \langle r^2 \rho u_r u_\theta \right \rangle = r^2 \overline{\rho} \widetilde{u_r} \widetilde{u_\theta} +{r^2\left \langle {\rho u_r^{\prime\prime} u_\theta^{\prime\prime}} \right \rangle},
\end{equation}
where $r^2 \overline{\rho} \widetilde{u_r} \widetilde{u_\theta}$ is the contribution from mean flow and  ${r^2\left \langle {\rho u_r^{\prime\prime} u_\theta^{\prime\prime}} \right \rangle}$ is the contribution from turbulent flow. The contribution from the mean flow is nearly negligible, with its maximum value being less than $1\%$ of the turbulence contribution (see inset in figure~\ref{fig:fluctuation}a). This suggests that the mean flow has minimal influence on the advection term in momentum transport. Since $r$ serves solely as a positional parameter, the advection term in momentum transport is completely determined by the Reynolds stress. Therefore, we can reinterpret momentum transport from the perspective of turbulence. 

Neutral droplets lead to an increase in Reynolds stress in the majority of the bulk region closer to the inner cylinder, whereas it decreases in the other half of the bulk region, thus accommodating the influence of the two-phase interface. However, the total Reynolds stress changes very little, and the drag enhancement is mainly due to the interfacial contribution.
For low-viscosity droplets, their lower viscosity weakens the inhibitory effect of the bulk region on turbulence compared to the continuous phase. However, the overall modulation of Reynolds stress is minimal in comparison to neutral droplets. Thus, the interfacial contribution remains the key factor in determining the drag enhancement.
In the case of light droplets, the aggregation of the light droplets near the inner cylinder results in a notable decrease in the upstream Reynolds stress (through $\rho$), primarily due to their density-related characteristics. Furthermore, the downstream Reynolds stress is correspondingly reduced (through $u_r^{\prime\prime} u_\theta^{\prime\prime}$) to ensure momentum conservation along the radial position, ultimately leading to drag reduction.
For low-viscosity light droplets, the alteration from $\xi_{\mu} = 1$ to $\xi_{\mu} = 1/4$ further reduces the momentum transport near the wall by decreasing the diffusion term (which is explicitly related to viscosity), thereby once again implicitly reducing the Reynolds stress in the bulk region (through $u_r^{\prime\prime} u_\theta^{\prime\prime}$) and causing a stronger drag reduction. In the studied parameter regime, neutral droplets and low-viscosity droplets increase the system’s drag due to the interfacial contribution, while light droplets and low-viscosity light droplets induce drag reduction by decreasing the Reynolds stress.

For the three components of TKE, we observe an overall decrease due to the presence of neutral droplets, indicating that the presence of the two-phase interface weakens the total turbulence intensity. 
Considering that Reynolds stress did not significantly decrease, it is likely that the presence of the two-phase interface leads to a transformation of turbulence or changes in the energy dissipation process within the system. Recent studies~\citep{perlekar2014spinodal,crialesi-esposito_rosti_chibbaro_brandt_2022} have demonstrated that the two-phase interface results in a decrease in energy for large-scale vortices and an increase in energy for small-scale vortices, suggesting an alteration in the energy cascade process. This alteration in the energy cascade process is likely responsible for the reduction observed in the three components of TKE, as the significant decrease in the total energy of large-scale vortices outweighs the increase in the total energy of small-scale vortices within the system. 
Furthermore, the slight increase in the three components of TKE for low-viscosity droplets compared to neutral droplets provides additional support for this conjecture. The slight increase indicates that reducing the viscosity of droplets weakens the reduction scale of TKE (or the alteration in the energy cascade process), aligning with the findings from~\cite{crialesi-esposito_rosti_chibbaro_brandt_2022}. From an alternative standpoint, the slight increase in the three components of TKE compared to neutral droplets can be attributed to the weakened inhibitory effect on turbulence caused by the decreased viscosity in the bulk region.

Light droplets cause a significant reduction in the three components of the TKE due to their explicit density dependence. Near the inner cylinder, this reduction primarily arises from the decreased density, whereas in the remaining region, it stems from the reduction in velocity fluctuations. Additionally, it is observed that low-viscosity light droplets result in a further reduction in the TKE components. This can be attributed to the fact that the low-viscosity droplets weaken momentum transport near the inner cylinder, subsequently causing a decrease in the energy available for transfer from the boundary layer to the bulk region.

Based on the analysis of turbulent fluctuations, we establish a connection between turbulence and the system’s drag through Reynolds stress. This approach provides more detailed insights into the effect of the two-phase interface, dispersed phase viscosity, and dispersed phase density on turbulence and the system’s drag.
In the case of two-phase flow with neutral droplets, the drag enhancement is primarily governed by the interfacial contribution. To maintain momentum transport conservation (a constant conserved quantity), the Reynolds stress is adjusted to match the presence of the two-phase interface, but the overall change is not significant.
Conversely, low-viscosity droplets will cause a slight increase in Reynolds stress within the bulk region since their low viscosity weakens the suppression of turbulence. However, the overall effect on the drag is not significant compared to the neutral droplets, and the interfacial contribution still dominates the drag enhancement. On the other hand, in the case of light droplets, their low density causes them to aggregate near the inner cylinder, resulting in a significant reduction of Reynolds stress in this region due to the explicit density dependence of Reynolds stress. The Reynolds stress in other regions decreases accordingly to ensure momentum transport conservation, ultimately leading to drag reduction.
When the low viscosity is introduced into light droplets, a more pronounced reduction in momentum transport occurs as the droplets predominantly aggregate near the inner cylinder, where viscosity plays a dominant role in momentum transport. Consequently, this leads to greater drag reduction.

To summarize, neutral droplets and low-viscosity droplets primarily contribute to drag enhancement through the interfacial contribution, while light droplets reduce the system's drag by reducing Reynolds stress. Furthermore, low-viscosity light droplets contribute to drag reduction mainly by reducing the diffusion contribution and Reynolds stress. It is important to note that viscosity plays different roles in different regions. Low viscosity near the wall results in a reduction in downstream Reynolds stress by decreasing upstream momentum transport, whereas low viscosity in the bulk region may lead to an increase in Reynolds stress.

\subsection{TKE budget analysis}

\begin{figure*}
  \centerline{\includegraphics[width=1\textwidth]{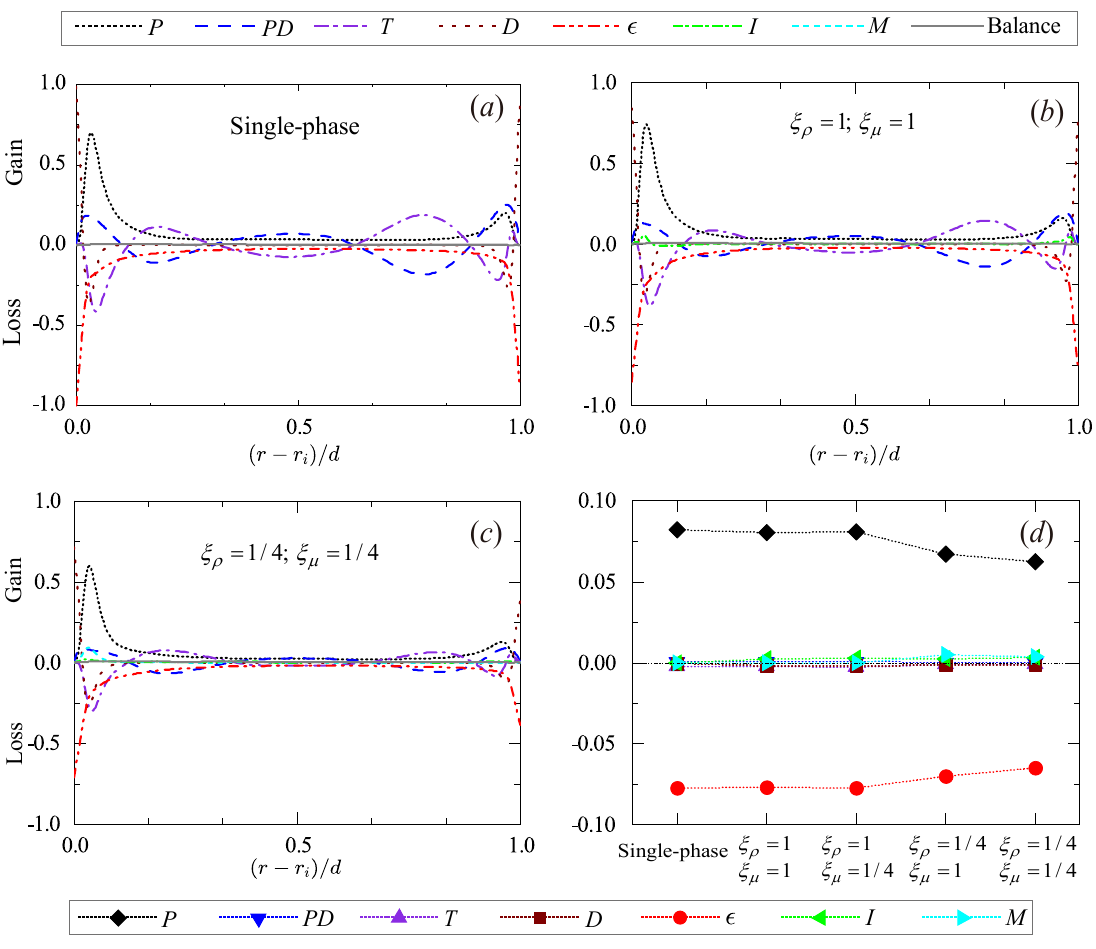}}%
  \caption{TEK Budget. Each term in the right hand of the equation~(\ref{equ:TKE2}) is shown for (\textit{a}) single-phase flow, (\textit{b}) two-phase flow with neutral droplets and (\textit{c}) two-phase flow with low-viscosity light droplets. (\textit{d}) The quantities are averaged in the radial direction to characterize the corresponding terms within the whole system. All the quantities are normalized by the $|\epsilon_{\varphi=0,i}|$, where $\epsilon_{\varphi=0,i}$ denotes the dissipation rate at the inner cylinder for sing-phase flow and the operator $|\cdot|$ gets the absolute value of the quantity.}
\label{fig:TKE tansport}
\end{figure*}

Considering the potential alterations in turbulence characteristics resulting from the introduction of a two-phase interface or changes in droplet fluid properties, it is not advisable to directly correlate the TKE with system’s drag. System’s drag imparts kinetic energy to the system, but a substantial fraction of this energy dissipates through turbulent mechanisms. Hence, examining the dissipation rate of TKE provides a valuable means to establish a significant linkage between system’s drag and turbulence properties. The transport equation of the TKE can be written in the form~\citep{besnard1992turbulence,wong2022analysis} 

\begin{equation}
 \begin{aligned}
  0
  =
  & \underbrace{ \left\langle{\rho\boldsymbol{u}^{\prime\prime} \boldsymbol{u}^{\prime\prime}}\right\rangle:(\nabla\widetilde{\boldsymbol{u}})}_P 
 \underbrace{-\nabla\cdot\left\langle{p^{\prime}\boldsymbol{u}^{\prime}}\right\rangle}_{PD}  \underbrace{-\nabla\cdot\left\langle{\rho\boldsymbol{u}^{\prime\prime}\boldsymbol{u}^{\prime\prime}\cdot\boldsymbol{u}^{\prime\prime}}\right\rangle/2}_T
  +\underbrace{\nabla\cdot \left\langle \boldsymbol{\tau}^{\prime}\cdot\boldsymbol{u}^{\prime}\right\rangle}_D
\underbrace{-\left\langle\boldsymbol{\tau}^{\prime}:\nabla\boldsymbol{u}^{\prime}\right\rangle}_\epsilon \\
&+
\underbrace{\left\langle\boldsymbol{u}^{\prime}\cdot \boldsymbol{f}^{\prime}\right\rangle}_I
+
\underbrace{\boldsymbol{a}\cdot(\nabla{\left\langle{p}\right\rangle -\nabla\cdot\left\langle\boldsymbol{\tau}\right\rangle})}_M,
  \end{aligned}
  \label{equ:TKE2}
\end{equation}
where $p^{\prime}=p-\left\langle{p}\right\rangle$, 
$\boldsymbol{\tau}^{\prime}=\boldsymbol{\tau}-\left\langle{\boldsymbol{\tau}}\right\rangle$, and $\boldsymbol{f}^{\prime}=\boldsymbol{f}-\left\langle{\boldsymbol{f}}\right\rangle$.
$P$ is the production term, $PD$ the pressure diffusion, $T$ the turbulence transport, $D$ the viscous diffusion, $\epsilon$ the dissipation term, $I$ interface term, and $M$ the additional production term due to the density difference between the dispersed and continuous phases. Figure~\ref{fig:TKE tansport} shows the distribution of each term in the TKE transport equation as a function of radial position. The $P$ denotes the energy transfer rate between the mean kinetic energy and the TKE and is consistently positive in our cases. It is mainly determined by Reynolds stress and azimuthal velocity gradient, i.e., $\left\langle \rho{u_r}^{\prime\prime}u_\theta^{\prime\prime}\right\rangle \partial_r \widetilde{u_\theta}$. The $\epsilon$ represents the dissipation rate of TKE and is always a negative value. 
\textcolor{black}{This value is directly related to the torque required to sustain the constant rotational velocity of the inner cylinder.} Note that the positive and negative values represent the gain and the loss in TKE, respectively. The pressure diffusion $PD$, turbulence transport $T$, and viscous diffusion $D$ often indicate the interchange or redistribution of the TKE. The balance term reflects any imbalance due to numerical effects and it is negligibly small. 

\textcolor{black}{For the two-phase flow with droplets, the presence of the two-phase interface introduces an interface term $I$ (see figure~\ref{fig:TKE tansport}b). By decreasing the dispersed phase density, an additional production term $M$ is observed mainly near the inner cylinder (see figure~\ref{fig:TKE tansport}c). By comparing with the single-phase case, we find that the pressure diffusion $PD$, turbulence transport $T$ and viscous diffusion $D$ are all reduced in two-phase cases, indicating that the presence of dispersed droplets weakens the interchange or redistribution of the TKE.} 

\begin{figure*}
  \centerline{\includegraphics[width=1\textwidth]{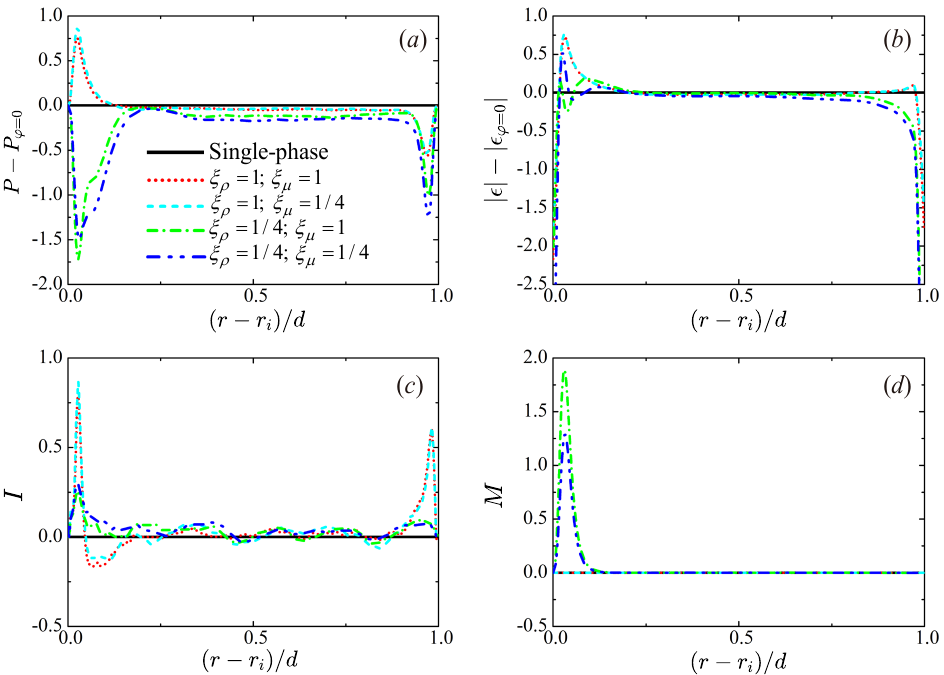}}%
  \caption{(\textit{a}) The production difference, (\textit{b}) the dissipation difference, (\textit{c}) the interface term, and (\textit{d}) the additional production term due to density difference between phases are shown as a function of radial position. All the quantities are normalized by the radial-averaged $|\epsilon_{\varphi=0}|$ to illustrate their magnitudes. The $P_{\varphi=0}$ and $\epsilon_{\varphi=0}$ denote the production term and dissipation term for the single-phase flow, respectively.}
\label{fig:TKE1}
\end{figure*}

To assess the overall impact of each term in equation~(\ref{equ:TKE2}) on the TKE transport, an analysis was conducted by globally averaging these terms, as illustrated in figure~\ref{fig:TKE tansport}(d). The findings indicate that TKE is mainly input into the system through the production term ($P$) and dissipated through the dissipation term ($\epsilon$). On the other hand, the effects of the pressure diffusion term ($PD$), turbulence transport term ($T$), and viscous diffusion term ($D$) are nearly negligible. Additionally, the interface term ($I$) and the additional production term ($M$), which are specific to two-phase flow, exhibit a small positive value, indicating that the two-phase interface and the density difference between the phases enhance the TKE transport. Consequently, the focus is on examining the modulation of the production term ($P$), dissipation term ($\epsilon$), interface term ($I$), and additional production term ($M$) with respect to their counterparts in single-phase flow, as depicted in figure~\ref{fig:TKE1}.

\textcolor{black}{As shown in figure~\ref{fig:TKE tansport}, the dissipation term is negative and indicates a loss of TKE. To facilitate observation and analysis, the absolute value of the dissipation term $|\epsilon|$ is adopted to characterize the magnitude of dissipation. The production difference ($P-P_{\varphi=0}$), dissipation difference ($|\epsilon|-|\epsilon_{\varphi=0}|$), interface term ($I$), and additional production term ($M$) are used to examine the effect of droplets on TKE transport, as depicted in figure~\ref{fig:TKE1}. A positive value indicates an enhancement effect on TKE transport, while a negative value suggests a weakening effect.} The interface term ($I$), specific to two-phase flow, is essentially positive, signifying its role in enhancing TKE transport. Moreover, for light droplets and low-viscosity light droplets, the additional production term ($M$) is predominantly positive near the inner cylinder, whereas it is close to zero in other regions. This term primarily arises from the interfacial structure adsorbed on the inner cylinder and contributes to the enhanced transport of TKE across the interface.

\begin{figure*}
  \centerline{\includegraphics[width=0.9\textwidth]{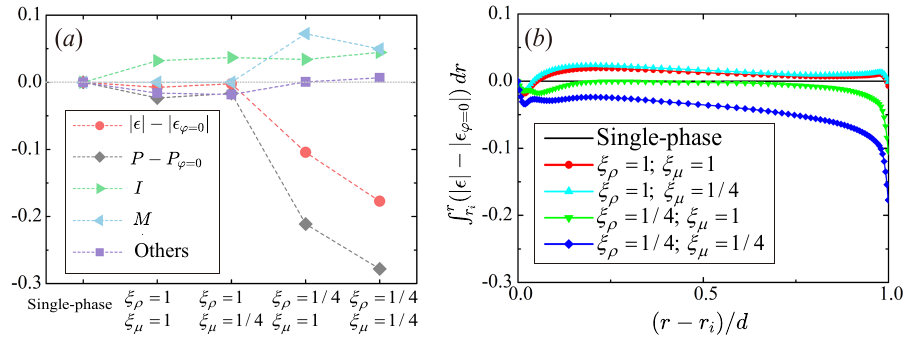}}%
  \caption{(\textit{a}) TKE budget analysis. The quantities are averaged in the radial direction to characterize the corresponding terms within the whole system. The ``Other'' denotes $(PD+T+D)-[PD+T+D]_{\varphi=0}$, where $[PD+T+D]_{\varphi=0}$ denotes the sum of $PD$, $T$, and $D$ for single-phase flow. (\textit{b}) The $|\epsilon|-|\epsilon_{\varphi=0}|$ is integrated within each cylindrical plane from the inner cylinder to a specific radial position. All the quantities are normalized by the radial-averaged $|\epsilon_{\varphi=0}|$ to illustrate their magnitudes.
}
\label{fig:new}
\end{figure*} 

For the two-phase flow with neutral droplets ($\xi_{\rho} = 1$ and $\xi_{\mu} = 1$), we observe net positive value in turbulence production near the inner cylinder, whereas net negative values are observed in the rest part of the domain. This indicates that the production term is enhanced near the inner cylinder, while the production term is weakened outside this region. \textcolor{black}{The dissipation difference shows a negative value in the region closest to the inner and outer cylinders and a positive value outside the closest region.} Correspondingly, the interface term primarily leads to an increase in TKE transport near the inner and outer cylinders. The locations where the dissipation difference is positive are approximately close to the regions where the interface term is positive, indicating that the interfacial tension enhances the dissipation near the cylinders. However, as depicted in figure~\ref{fig:new}(a), when considering the global perspective, although the two-phase interface contributes to the TKE transport through the interfacial tension, it also results in a decrease in the production term. This ultimately leads to a slight reduction in the dissipation term, indicating a decrease in the amount of power required to sustain turbulence. Therefore, the drag enhancement induced by the neutral droplets weakly depends on the TKE transport. Furthermore, the moderating effect of low-viscosity droplets on TKE transport is similar to that of neutral droplets. This similarity suggests that the drag enhancement caused by low-viscosity droplets is also weakly dependent on the TKE transport.

In the case of two-phase flow with light droplets ($\xi_{\rho} = 1/4$ and $\xi_{\mu} = 1$), the presence of these droplets weakens turbulence production throughout the entire domain. This weakening is attributed to the reduced Reynolds stress, which leads to a significant reduction in the total turbulence production, as depicted in figure~\ref{fig:TKE1}(a) and figure~\ref{fig:new}(a). 
From a global perspective, there is a negligible modulation effect on the sum of pressure diffusion, turbulence transport, and viscous diffusion. By isolating the effects of the interface term and the additional production term resulting from density differences, it is observed that a total of $10.4\%$ of the turbulence dissipation is reduced. This reduction indicates that the power required to sustain turbulence in the system has decreased by this amount, ultimately resulting in a drag reduction. It is worth noting that the reduction in the total dissipation rate is primarily due to the decrease in the dissipation rate near the outer cylinder, as shown in figure~\ref{fig:TKE1}(b) and figure~\ref{fig:new}(b). This is because the density contrast between the phases boosts TKE transport near the inner cylinder, thus weakening the reduction in the total dissipation rate there.

\textcolor{black}{In the case of two-phase flow with low-viscosity light droplets \textcolor{black}{($\xi_{\rho} = 1/4$ and $\xi_{\mu} = 1/4$)}, a more pronounced reduction in the dissipation rate is observed. Specifically, a total of $17.7\%$ of the power required to sustain turbulence in the system has been decreased. The transition from light droplets to low-viscosity light droplets leads to a further decrease in the Reynolds stress, resulting in a more significant weakening of turbulence production. Moreover, the reduction in droplet viscosity diminishes the impact of the additional production term while having minimal effects on the other terms. As a result, a larger reduction in the turbulent dissipation rate is achieved, leading to a more substantial drag reduction.} 

\textcolor{black}{Based on the aforementioned analyses, it is evident that two-phase flow introduces an interface term and an additional production term arising from the density difference between the two phases. 
While the interfacial tension essentially enhances the TKE transport, the drag enhancement is not strongly correlated with TKE transport. This mechanism holds for both neutral droplets and low-viscosity droplets, as observed in our examination.
Light droplets, on the other hand, cause a reduction in the production term mainly by reducing the Reynolds stress. However, both the density difference between the two phases and the interface between the two phases lead to an increase in TKE transport. Consequently, the reduction in the turbulent dissipation rate is mainly attributed to the decrease in the turbulence production term, which in turn causes drag reduction.
The production term is further reduced in the case of low-viscous light droplets due to their greater reduction in Reynolds stress. In addition, the reduction in viscosity weakens the additional production term. As a result, a greater drag reduction is achieved.
}

\section{Conclusions}

\textcolor{black}{In this study, we have examined how the dispersed phase influences system’s drag and turbulence properties in a two-phase fluid-fluid Taylor-Couette system operating at a system Reynolds number of $6\times10^3$ \textcolor{black}{and a system Weber number of 10}. To achieve this, we employ the interface-resolved volume-of-fluid method, which enables the resolution of two-phase flows through a single-equation formulation. This approach allows us to conduct operations similar to single-phase flow and facilitates the exploration of how the dispersed phase influences the statistical properties of turbulence.
}

Through our analysis of momentum transport and turbulent fluctuations, we establish a connection between turbulence statistics and the system’s drag using Reynolds stress. This provides detailed insights into the impact of the two-phase interface, dispersed phase viscosity, and dispersed phase density on turbulence and the system’s drag.
In the case of two-phase flow with neutral droplets, drag enhancement is primarily governed by the interfacial contribution. The Reynolds stress is modulated to account for the influence of the two-phase interface, but the overall change is not significant.
Lowering the droplets' viscosity will cause a slight increase in Reynolds stress, as their low viscosity weakens the suppression of turbulence within the bulk region. However, the overall effect on drag is not significant compared to neutral droplets, and the interfacial contribution still dominates drag enhancement.
On the other hand, in the case of light droplets, their low density causes them to aggregate near the inner cylinder, resulting in a significant reduction in Reynolds stress in this region due to the explicit density dependence of Reynolds stress. The Reynolds stress in other regions decreases accordingly to ensure momentum transport conservation, ultimately leading to drag reduction.
When low viscosity is introduced into light droplets, a more pronounced reduction in Reynolds stress occurs, as the droplets predominantly aggregate near the inner cylinder, where viscosity plays a dominant role in momentum transport. Consequently, this leads to greater drag reduction.
In summary, neutral droplets and low-viscosity droplets primarily contribute to drag enhancement through the interfacial contribution, while light droplets reduce the system’s drag by explicitly reducing Reynolds stress. Furthermore, low-viscosity light droplets contribute to drag reduction by explicitly and implicitly reducing Reynolds stress.

Furthermore, we have investigated the effect of the dispersed phase on the transport of turbulent kinetic energy (TKE). In two-phase flow, an interface term arises due to interfacial tension, potentially accompanied by an additional production term resulting from the density difference between the two phases.
While interfacial tension primarily enhances TKE transport, the correlation between drag enhancement and TKE transport is not strong, as the two-phase interface also weakens turbulence production. This holds true for both neutral droplets and low-viscosity droplets.
On the other hand, light droplets reduce the production term primarily by lowering Reynolds stress. However, the density contrast between the two phases enhances TKE transport near the inner wall. Consequently, the reduction in the turbulence dissipation mainly stems from the decrease in the production term, leading to drag reduction.
In the case of low-viscosity light droplets, the production term is further reduced due to a greater decrease in Reynolds stress. Additionally, the decrease in viscosity weakens the additional production term. This results in a more significant reduction in the turbulence dissipation, leading to stronger drag reduction.

\textcolor{black}{From the perspective of turbulence, we find that the Reynolds stress plays a key role in drag modulation. It participates in momentum transport in the form of an advection contribution and participates in TKE transport as an important component of the turbulence production term. 
From a drag reduction perspective, we emphasize the critical role that density \textcolor{black}{ratio} plays in this process.  
By reducing the density of the dispersed phase, the aggregation of the dispersed phase near the inner cylinder is promoted, which amplifies the influence of lower density on Reynolds stress, thereby causing significant drag reduction at a very small volume fraction of the dispersed phase. 
Furthermore, it enables greater drag reduction by promoting the transportation of low-viscosity droplets towards the near-wall region where viscosity plays a dominant role in momentum transport. }

\backsection[Funding]
{This work is financially supported by the National Natural Science Foundation of China under Grant Nos. 11988102, 22478421, 12402299 and 12402298, the New Cornerstone Science Foundation through the New Cornerstone Investigator Program and the XPLORER PRIZE, and the Science Foundation of China University of Petroleum, Beijing (No. 2462024YJRC008).}

\backsection[Declaration of Interests]{The authors report no conflict of interest.}

\backsection[Author ORCIDs]{
\\Jinghong Su https://orcid.org/0000-0003-1104-6015;
\\Cheng Wang https://orcid.org/0000-0002-6470-7289;
\\Yi-Bao Zhang https://orcid.org/0000-0002-4819-0558;
\\Fan Xu https://orcid.org/0009-0004-3324-7859;
\\Junwu Wang https://orcid.org/0000-0003-3988-1477; 
\\Chao Sun https://orcid.org/0000-0002-0930-6343.}

\appendix
\section{Conservation of momentum transport}\label{appA}
\textcolor{black}{In TC flow, $J^\omega$ should be constant in the radial direction, but numerically
it does show some dependence on the radial direction. Because of numerical errors, $J^\omega$ will deviate slightly from being constant. To quantify this difference,~\cite{zhu2016direct} defined
\begin{equation}
\Delta_J = \frac{{\rm max}(J^{\omega}(r))-{\rm min}(J^{\omega}(r))}{\langle J^{\omega}(r) \rangle_r},
\end{equation}
where the  operator $\langle \cdot \rangle_r$ denotes average in the radial direction.
As illustrated by~\cite{ostilla2013optimal} and ~\cite{zhu2016direct}, $\Delta_J \leq 1\%$ is the very strict requirement for the single-phase case. We make sure that the single-phase case meets this criterion. For the two-phase cases, due to the additional errors introduced by the two-phase interface and the disparities in liquid properties between the two phases, $\Delta_J$ will be slightly larger than $1\%$ for the two-phase flow as shown in table~\ref{tab:J}.}

\begin{table}
    	\centering
     \setlength\tabcolsep{20pt}
		\begin{tabular}{ccccc}
			\textrm{$\varphi$}&
   		\textrm{$\rm {Re}$}&
			\textrm{$\xi_{\rho}=\rho_d /\rho_f$}&
			\textrm{$\xi_{\mu}= \mu_d /\mu_f$}&
			\multicolumn{1}{c}{$\Delta_J \times 100\% $}\\[6pt]

                $0  $  &  $4389$  & ---	    & ---	 		 & $0.33\%$		              \\[2pt]
                $0  $  &  $5067$  & ---		& ---	 		 & $0.68\%$		              \\[2pt]
			$0  $  &  $6000$  & ---		& ---	    	 & $0.64\%$		              \\[2pt]
			$5\%$  &  $6000$  & $1$	    & $1$	 	     & $0.71\%$ 	      \\[2pt]
   	      $5\%$  &  $6000$  & $1$	  & $1/4$	       & $0.73\%$ 	      \\[2pt]
			$5\%$  &  $6000$  &$1/4$    & $1$	         & $1.08\%$	       \\[2pt]
			$5\%$  &  $6000$  & $1/4$   & $1/4$	 	     & $1.26\%$ 	       \\[2pt]
			$10\%$ &  $6000$  & $1$	    & $1$	 	     & $0.66\%$  	       \\[2pt]
      	  $10\%$ &  $6000$  & $1$	  & $1/4$	       & $0.87\%$ 	      \\[2pt]
			$10\%$ &  $6000$  & $1/4$   & $1$	 	     & $1.05\%$ 	       \\[2pt]
			$10\%$ &  $6000$  & $1/4$   & $1/4$	 	     & $1.32\%$ 	       \\[2pt]
		\end{tabular}
  \caption{Radial deviation of the $J^{\omega}$. }
  \label{tab:J}
\end{table}

\section{Data validation and resolution test}\label{appB}

\textcolor{black}{To validate the accuracy of our simulations, we have simulated two cases with the Taylor number being $3.90 \times 10^6$ ($Re=1600$) and $9.52 \times 10^6$ ($Re=2500$) and validated our results through comparisons with those from~\cite{ostilla2013optimal} in our previous work~\citep{su2024numerical}. Because the $Re=6000$ studied in this work is not included in the work of~\cite{ostilla2013optimal}, we additionally simulate a single phase case at $Re = 5600$ and compare the dimensionless conserved quantity $Nu_\omega=J^{\omega}/J^{\omega}_{lam}$ with that from~\cite{ostilla2013optimal}, where $J^{\omega}_{lam}$ corresponding to the case when the flow is fully laminar. In our work, the minimum flow geometry with a rotational symmetry of six ($n_{sym} = 6$, i.e., the azimuthal angle of the simulated domain is $\pi/3$) and an aspect ratio of $\Gamma=L/d=2\pi/3$ is selected to reduce the computational cost while not affecting the results, which has been verified by previous studies~\citep{brauckmann2013direct,ostilla2015effects}. We represent the $Nu_{\omega}$ from~\cite{ostilla2013optimal} and OpenFOAM as $Nu_{\omega,Ot}$ and $Nu_{\omega,Op}$, respectively, as shown in table~\ref{tab:validation}. The deviation is obtained by ($Nu_{\omega,Op}/Nu_{\omega,Ot}$)-1. A deviation of $-1.18\%$ is found, indicating that our simulation is sufficient to capture the flow field information. }

\begin{table}
    	\centering
     \setlength\tabcolsep{6pt}
		\begin{tabular}{cccccccc}
			\textrm{}&
			\textrm{$Re$}&
                \textrm{$Ta$}&
                \textrm{$n_{sym}$}&
                 \textrm{$\Gamma$}&
                 \textrm{$N_\theta \times N_r \times N_z$}&
                 \textrm{$Nu_\omega$}&
			\multicolumn{1}{c}{Deviation}\\[6pt]
			Ostilla et al.    & 5600	& $4.77\times10^7$ & $1$	    & $2\pi$		 & $800\times400\times400$	    & 8.78178     & --- \\[2pt]
			Our simulation          & 5600 	& $4.77\times10^7$ & $6$	    & $2\pi/3$       & $200\times256\times192$	    & 8.67926     & $-1.18\%$\\[2pt]
		\end{tabular}
     	\caption{\label{tab:validation}%
		\textcolor{black}{Validation of the calculation results for the single-phase flow at Re = 5600.}
	}
\end{table}

\begin{figure}
	\centering
	\includegraphics[width=0.6\linewidth]{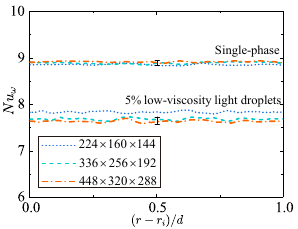}
	\caption{Radial dependence of $Nu_\omega$ for three different grid resolutions. An error bar indicating a $1\%$ error is provided for the single-phase case and an error bar indicating a $2\%$ error is provided for the two-phase flow with low-viscosity light droplets for reference.
	}
	\label{fig:S01}
\end{figure}

\textcolor{black}{To obtain reliable numerical results, the grid's spatial resolutions have to be sufficient. The resolution test is conducted for the single-phase flow and the two-phase flow with $5\%$ low-viscosity light droplets. 
A roughly resolved case ($N_\theta \times N_r \times N_z = 224\times160\times144$) with the maximum grid spacing being about 2.02$\eta_k$, a reasonably resolved case ($N_\theta \times N_r \times N_z = 336\times256\times192$) with the maximum grid spacing being about 1.5$\eta_k$ and a well-resolved case ($N_\theta \times N_r \times N_z = 448\times320\times288$) with the maximum grid spacing being about $1.01\eta_k$ are considered for resolution test as depicted in figure~\ref{fig:S01}. For the grid with $448\times320\times288$, most of the grid spacing is less than $\eta_k$. 
We have shown the three types of grid resolution in table~\ref{tab:mesh} and compared them to the grid used by \cite{chouippe2014numerical} to investigate the bubble dispersion in turbulent Taylor-Couette flow at Re = 5000. All three grid resolutions have a finer grid spacing in the radial and azimuthal directions than those used by \cite{chouippe2014numerical}. For the grid spacing in the axial direction, the two finer grid resolutions used in this work have a similar scale to the grid used by \cite{chouippe2014numerical}. 
It is observed that the grid with $224\times160\times144$ can roughly resolve the single-phase case (see figure~\ref{fig:S01}). For the two-phase case with $5\%$ low-viscosity light droplets, the grid with $224\times160\times144$ shows a distinguishable difference of $Nu_\omega$ compared to the two finer grid resolutions. For the two finer grids, both the $Nu_\omega$ lie within $1\%$ error bar for the single-phase cases.  Because the two-phase interface as well as the disparities in liquid properties between the two phases could introduce additional numerical errors, we provide an error bar of $2\%$ as a reference. Here, $\Delta_J =1.26\%$ for $336\times256\times192$ and $\Delta_J =1.35\%$ for $448\times320\times288$. Besides, there is a deviation of $0.67\%$ for $\langle J^{\omega}(r) \rangle_r$ when comparing the results from the two finer grids. Therefore, the adopted spatial resolution $N_\theta \times N_r \times N_z = 336\times256\times192$ is sufficient to obtain reliable results for both single-phase and two-phase cases studied. }

\begin{table}
    	\centering
     \setlength\tabcolsep{4pt}
		\begin{tabular}{cccccccccc}
			\textrm{Grid}&
			\textrm{$Re$}&
                \textrm{$\eta$}&
                \textrm{$\Gamma$}&
                \textrm{$n_{sym}$}&
                 \textrm{$N_\theta \times N_r \times N_z$}&
                 \textrm{$\Delta r^+$}&
                 \textrm{$r_i\Delta \theta^+$}&
			\multicolumn{1}{c}{$\Delta z^+$}\\[6pt]
      Chouippe et al.     & 5000     & $0.72$   & $2.09$		& $1$     & $400\times100\times200$	      & $0.65 \sim 4.48$    & $10$  & $2.66$\\[2pt]
    		Grid1     & 6000 	 & $0.714$	& $2\pi/3$    & $6$     & $224\times160\times144$	    & $0.55 \sim 4.38$    & $3.46$  & $4.31$\\[2pt]
   			Grid2     & 6000 	 & $0.714$	& $2\pi/3$    & $6$     & $336\times256\times192$	    & $0.34 \sim 2.74$    & $2.31$  & $3.23$\\[2pt]
                Grid3     & 6000 	 & $0.714$	& $2\pi/3$    & $6$     & $448\times320\times288$	    & $0.27 \sim 2.19$    & $1.73$  & $2.15$\\[2pt]
		\end{tabular}
     	\caption{\label{tab:mesh}%
		\textcolor{black}{Grid resolutions. The grid spacing is shown in units of the viscous length scale $\delta_{v}$ for single-phase case.}
	}
\end{table}

\begin{figure*}
\centerline{\includegraphics[width=1\textwidth]{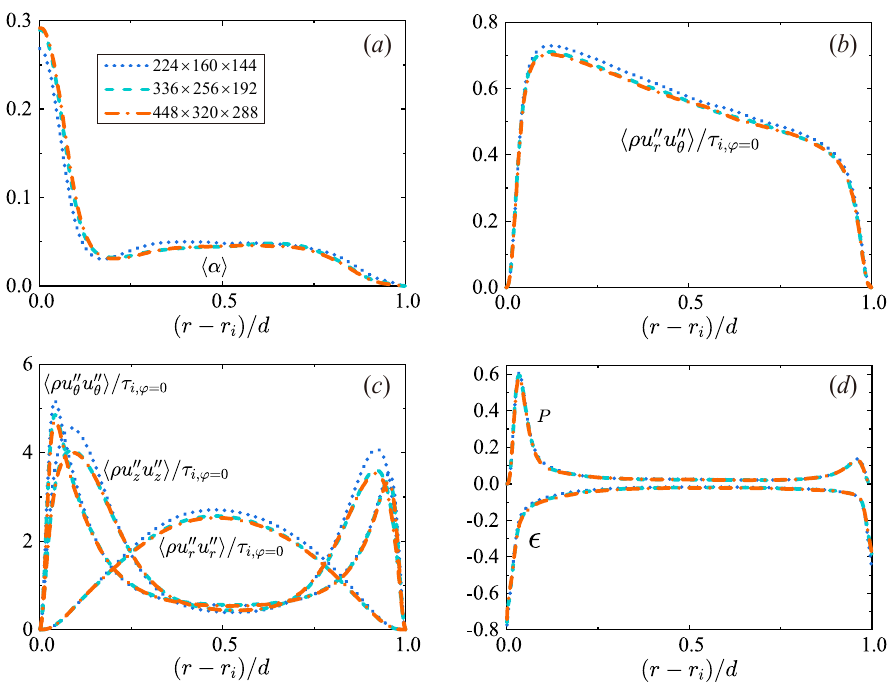}}%
  \caption{ Radial dependence of (\textit{a})  phase fraction distribution, (\textit{b})  Favre-averaged Reynolds stress, (\textit{c}) three components of the TKE and (\textit{d})  production and dissipation terms of the TKE transport for three different grid resolutions for the $5\%$ low-viscosity light droplets case. The terms in (\textit{d}) are normalized by the $|\epsilon_{\varphi=0,i}|$.
 }
\label{fig:mesh2}
\end{figure*}

\textcolor{black}{The resolution sensitivity of the statistics discussed in this work is also conducted for the two-phase case with $5\%$ low-viscosity light droplets. Considering that the distribution of the phase fraction may be dependent on the grid resolution, the distribution of phase fraction for three different grid resolutions is shown in figure~\ref{fig:mesh2}(a). 
Besides, given that turbulent statistics are higher-order and more challenging to get converged results compared to the lower-order results such as mean velocity profiles, we additionally show Favre-averaged Reynolds stress, three components of the TKE as well as production term and dissipation term of the TKE transport for three different grid resolutions as depicted in figure~\ref{fig:mesh2}(b-d). All results are nearly identical for the two finer grid resolutions, indicating that the statistics discussed in this work are qualitatively converged. It is observed that the production term and dissipation term of the TKE transport are visually nearly identical for all three grid resolutions, which is due to the large difference between the maximum and minimum values in the plot.}

\begin{figure*}
\centerline{\includegraphics[width=1\textwidth]{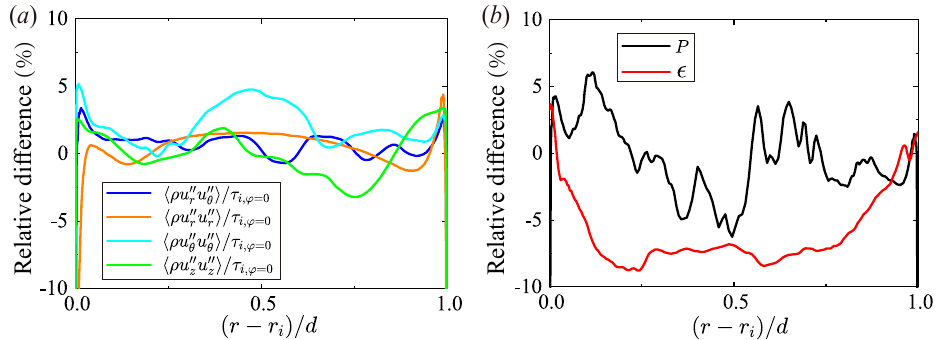}}%
  \caption{(\textit{a}) Relative difference of Reynolds stress and three components of TKE for the two finer grid resolutions. (\textit{b}) Relative difference of the production and dissipation terms of the TKE transport for the two finer grid resolutions. 
 }
\label{fig:relative difference}
\end{figure*}

\textcolor{black}{To show the effect of the two finer grid resolutions in more detail, we plot the relative differences as a function of the radial position (see figure~\ref{fig:relative difference}). The relative differences are calculated by $Q_1/Q_2-1$, where $Q_1$ represents the quantity for the grid with $336\times256\times192$ and $Q_2$ represents the quantity for the grid with $448\times320\times288$. It's observed that near the two cylinders, the magnitude of relative difference could be larger than $10\%$ when the quantity $Q_2$ is close to zero. This is acceptable considering that the absolute difference would be very small.
In the bulk region, the magnitude of the relative difference for the Reynolds stress is within $2\%$, which has a similar scale as that of the conserved quantity since the conserved quantity in the bulk region is primarily determined by the Reynolds stress.
For the three components of TKE, the magnitude of relative differences is within $5\%$ in the bulk region. 
Because the production and dissipation terms are high-order quantities, the magnitude of relative difference could be up to nearly $10\%$ in the bulk region. Since the magnitude of $P$ and $\epsilon$ in the bulk region are much smaller than those near the inner cylinder, the relative difference is difficult to distinguish visually in figure~\ref{fig:mesh2}(d). In two-phase flow simulations, the relative differences in figure~\ref{fig:relative difference} are acceptable for judging that the grid with $336\times256\times192$ is sufficient to provide reliable turbulent statistics.}

\section{Mean streamwise velocity profiles}\label{appC}

\begin{figure*}
\centerline{\includegraphics[width=0.5\textwidth]{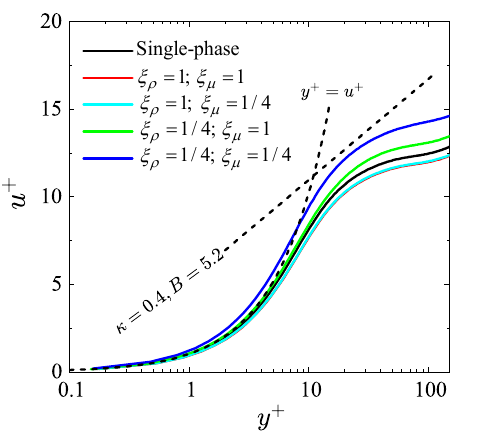}}%
  \caption{Mean azimuthal velocity profiles near the inner cylinder. $u^+=(u_i-\langle u_{\theta} \rangle)/u_\tau$ is the velocity difference from the inner cylinder normalized by the friction velocity $u_\tau$. $y^+=(r-r_i)/\delta_{v}$ is the distance from the inner cylinder in units of the viscous length scale. The dotted lines show the linear relation $u^+=y^+$ and the logarithmic law $u^+=(1/\kappa)\mathrm{ln}y^++B$ with the typical values $\kappa=0.4$ and $B=5.2$~\citep{huisman2013logarithmic}.
 }
\label{fig:u+}
\end{figure*}

In figure~\ref{fig:u+}, we plot the mean azimuthal velocity profiles normalized by the frictional velocity, $u^+=(u_i-\langle u_{\theta} \rangle)/u_\tau$, versus the wall distance $y^+=(r-r_i)/\delta_{v}$.
\textcolor{black}{Given the differences in fluid properties in the cases with light droplets and low-viscosity light droplets and uneven distribution of the droplets, it is difficult to obtain proper density and viscosity to calculate friction velocity $u_\tau$ and viscous length scale $\delta_\nu$.
Different choices of equivalent density and equivalent viscosity will cause differences in results. For ease of analysis, we have taken the density and viscosity of the continuous phase to calculate the friction velocity and viscous length scale, i.e., $u_\tau=\sqrt{\tau_{i}/\rho_f}$ and $\delta_\nu = \nu_f/u_\tau$. The shear stress at the inner cylinder is obtained by $\tau_{i} = \left \langle \mu r_i \partial _r \omega \right \rangle$ for different cases, where $\mu$ is the dynamic viscosity of the combined phase. 
For the single phase case, $u^+$ follows the linear relation $u^+=y^+$ well in the viscous sublayer ($y^+<5$). 
$u^+$ at $y^+>30$ does not exhibit a clear logarithmic shape due to the small $Re$ in this study \citep{huisman2013logarithmic}. Moreover, the velocity profile shows an obvious discrepancy from the classical logarithmic law profile with the typical values $\kappa=0.4$ and $B=5.2$, which is because the boundary layer is not yet fully developed and the zero pressure gradient boundary layer is not satisfied in the Taylor-Couette flow~\citep{ostilla2014boundary}. For the cases with neutral droplets and low-viscosity droplets, $u^+$ shifts slightly downward in the buffer layer and above, which has also been observed in the case of drag enhancement caused by ﬁnite-size particles in plane-Couette ﬂow~\citep{wang_jiang_sun_2023}. The downward shift can be interpreted as a thinning of the buffer layer.}  Differently, $u^+$ shifts slightly upwards in the buffer layer and above for the case with light droplets, which is consistent with the case of drag reduction caused by the polymer in turbulent channel flow~\citep{li2006influence}. The upward shift can be interpreted as a thickening of the buffer layer. For the case with low-viscosity light droplets, $u^+$ shifts upward again due to their stronger drag reduction effect than that of the light droplets. In addition, due to the effect of the low viscosity of the droplets within the viscous sublayer, $u^+=y^+$ is not valid anymore.

\section{The effect of effective Reynolds numbers}\label{appD}

\begin{figure*}
\centerline{\includegraphics[width=0.5\textwidth]{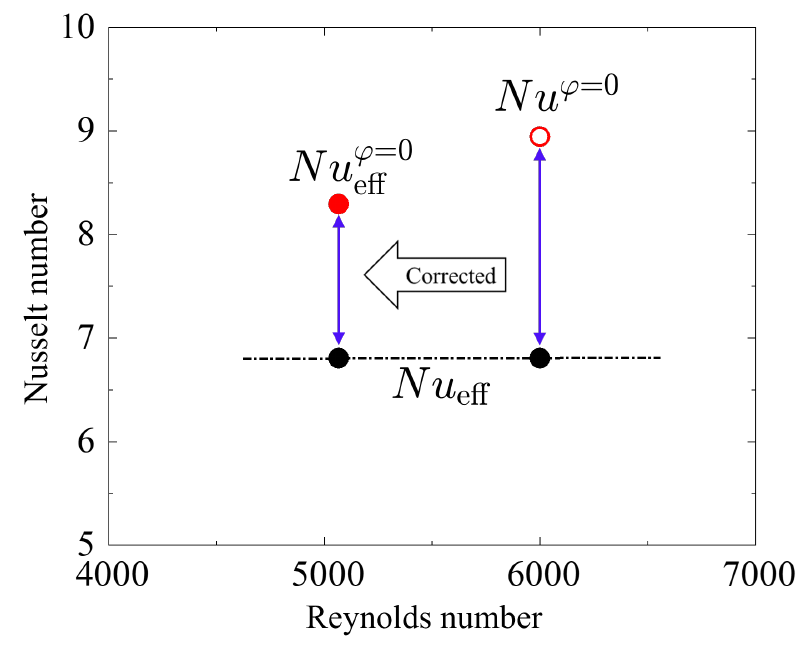}}%
  \caption{The correction of the drag modulation characterized by the effective Nusselt number for the case with $5\%$ low-viscosity light droplets.}
\label{fig:S05}
\end{figure*}

\textcolor{black}{In the case of light droplets and low-viscosity light droplets, one may wonder that the drag reduction effect is due to the difference in effective Reynolds numbers. Therefore, it is necessary to correct the Reynolds number to exclude the possibility of drag reduction due to different effective Reynolds numbers. The effective density can be approximated using $\rho_\varphi= (1-\varphi)\rho_f+ \varphi \rho_d$.
As for the effective viscosity, we note that there is no model available to estimate the effective viscosity of a liquid-liquid two-phase system. Our previous experimental results found that the effective viscosity increases with the volume fraction of the dispersed phase \citep{yi2021global}. At low volume fractions and low Reynolds numbers, the effective viscosity is very close to the model of Krieger and Dougherty (KD) \citep{krieger1959mechanism}. In our study, the volume fraction is in the range of $0\leq \varphi \leq 10\%$, and ${\rm Re}=6000$. We therefore employ the KD model to estimate the effective dynamic viscosity, i.e., $\mu_{\varphi} = \mu_f(1 - \varphi/\varphi_m)^{-2.5\varphi_m}$, where $\varphi_m = 0.58$. 
The Reynolds number corrected by the effective density and effective viscosity can be defined as ${\rm Re_{\rm eff}}=\rho_\varphi u_i d/\mu_\varphi$, which is 5067 and 4389 for $\varphi=5\%$ and $\varphi=10\%$, respectively.
The effective Nusselt number can be defined as $Nu_{\rm eff}=T/(4 \pi \mu_{\varphi} L r_i^2 r_o^2 \omega_i / (r_o^2-r_i^2))$ and the uncorrected drag modulation based on the effective Nusselt number can be written
as $Nu_{\rm eff}/Nu^{\varphi=0} -1 = (T/T_{\varphi=0})(\mu_f / \mu_\varphi)-1$ as shown in table~\ref{tab:drag correct}, where $Nu^{\varphi=0}=T_{\varphi=0}/(4 \pi \mu_f L r_i^2 r_o^2 \omega_i / (r_o^2-r_i^2))$ is the Nusselt number for the single-phase case at ${\rm Re}=6000$. Since the effective Reynolds number in the two-phase case is different from the single-phase case, we have used $T/T_{\varphi=0}-1$ rather than $Nu_{\rm eff}/Nu^{\varphi=0} -1$ to characterize the drag modulation as shown in table~\ref{tab:drag}.
The corrected drag modulation based on the effective Nusselt number can be obtained by $Nu_{\rm eff}/Nu_{\rm eff}^{\varphi=0} -1$, where $Nu_{\rm eff}^{\varphi=0}$ is the Nusselt number for the single-phase case at the effective Reynolds number. To obtain the corrected drag modulation, we additionally simulate two single-phase cases at these two effective Reynolds numbers (${\rm Re_{\rm eff}}=5067$ and 4389) and obtain the corrected drag modulation by $Nu_{\rm eff}/Nu_{\rm eff}^{\varphi=0} -1$. 
See figure \ref{fig:S05} for an illustration of the correction procedure, and the drag modulation before and after the correction is shown in the table \ref{tab:drag correct}. 
Although the magnitude of drag reduction is reduced when the different Reynolds numbers between the two-phase and single-phase cases are taken into account, the drag reduction effect by the light droplets and low-viscosity light droplets is still significant. Therefore, we conclude that the drag reduction can not be attributed to the discrepancy in the effective Reynolds number.
}

\begin{table}
    	\centering
     \setlength\tabcolsep{6pt}
		\begin{tabular}{ccccccc}
			\textrm{$\varphi$}&
			\textrm{$\xi_{\rho}=\rho_d /\rho_f$}&
			\textrm{$\xi_{\mu}= \mu_d /\mu_f$}&
      	    \textrm{$\rm Re$}&
   			\textrm{$Nu_{\rm eff}/Nu^{\varphi=0}-1$}&
   			\textrm{${\rm Re}_{\rm eff}$}&
			\multicolumn{1}{c}{$Nu_{\rm eff} / Nu_{\rm eff}^{\varphi=0} -1$}\\[6pt]
			$5\%$ & $1/4$ & $1$	     & $6000$   & $-16.61\%$   & $5067$   & $-10.05\%$   \\[2pt]
			$5\%$ & $1/4$ & $1/4$    & $6000$   & $-23.92\%$  & $5067$   & $-17.92\%$    \\[2pt]
			$10\%$ & $1/4$ & $1$	 & $6000$   & $-36.37\%$  & $4389$    & $-25.97\%$   \\[2pt]
			$10\%$ & $1/4$ & $1/4$	 & $6000$   & $-46.62\%$  & $4389$    & $-37.89\%$  \\[2pt]
		\end{tabular}
  \caption{\label{tab:drag correct} Drag modulation of light droplets and low-viscosity light droplets.}
\end{table}

\bibliographystyle{jfm}
\bibliography{jfm-instructions}

\begin{thebibliography}{56}
\expandafter\ifx\csname natexlab\endcsname\relax\def\natexlab#1{#1}\fi
\def\au#1{#1} \def\ed#1{#1} \def\yr#1{#1}\def\at#1{#1}\def\jt#1{\textit{#1}}
  \def\bt#1{#1}\def\bvol#1{\textbf{#1}} \def\vol#1{#1} \def\pg#1{#1}
  \def\publ#1{#1}\def\arxiv#1{#1}\def\org#1{#1}\def\st#1{\textit{#1}}

\bibitem[Assen {\em et~al.\/}(2022)Assen, Ng, Will, Stevens, Lohse \&
  Verzicco]{assen2022strong}
{\sc \au{Assen, Martin~PA}, \au{Ng, Chong~Shen}, \au{Will, Jelle~B},
  \au{Stevens, Richard~JAM}, \au{Lohse, Detlef} \& \au{Verzicco, Roberto}}
  \yr{2022}  \at{Strong alignment of prolate ellipsoids in {T}aylor--{C}ouette
  flow}.  \jt{J. Fluid Mech.}  \bvol{935},  \pg{A7}.

\bibitem[Bakhuis {\em et~al.\/}(2021)Bakhuis, Ezeta, Bullee, Marin, Lohse, Sun
  \& Huisman]{bakhuis2021catastrophic}
{\sc \au{Bakhuis, Dennis}, \au{Ezeta, Rodrigo}, \au{Bullee, Pim~A}, \au{Marin,
  Alvaro}, \au{Lohse, Detlef}, \au{Sun, Chao} \& \au{Huisman, Sander~G}}
  \yr{2021}  \at{{Catastrophic phase inversion in high-Reynolds-number
  turbulent Taylor-Couette flow}}.  \jt{Phys. Rev. Lett.}  \bvol{126}~(6),
  \pg{064501}.

\bibitem[Bakhuis {\em et~al.\/}(2018)Bakhuis, Verschoof, Mathai, Huisman, Lohse
  \& Sun]{bakhuis2018finite}
{\sc \au{Bakhuis, Dennis}, \au{Verschoof, Ruben~A}, \au{Mathai, Varghese},
  \au{Huisman, Sander~G}, \au{Lohse, Detlef} \& \au{Sun, Chao}} \yr{2018}
  \at{{Finite-sized rigid spheres in turbulent Taylor--Couette flow: effect on
  the overall drag}}.  \jt{J. Fluid Mech.}  \bvol{850},  \pg{246--261}.

\bibitem[Besnard {\em et~al.\/}(1992)Besnard, Harlow, Rauenzahn \&
  Zemach]{besnard1992turbulence}
{\sc \au{Besnard, Didier}, \au{Harlow, Francis~H}, \au{Rauenzahn, Rick~M} \&
  \au{Zemach, Charles}} \yr{1992}  \bt{{Turbulence transport equations for
  variable-density turbulence and their relationship to two-field models}}.
  {\em Tech. Rep.\/}.  \org{Los Alamos National Lab.(LANL), Los Alamos, NM
  (United States)}.

\bibitem[Brackbill {\em et~al.\/}(1992)Brackbill, Kothe \&
  Zemach]{brackbill1992continuum}
{\sc \au{Brackbill, Jeremiah~U}, \au{Kothe, Douglas~B} \& \au{Zemach, Charles}}
  \yr{1992}  \at{{A continuum method for modeling surface tension}}.  \jt{J.
  Comput. Phys.}  \bvol{100}~(2),  \pg{335--354}.

\bibitem[Brauckmann \& Eckhardt(2013)]{brauckmann2013direct}
{\sc \au{Brauckmann, Hannes~J} \& \au{Eckhardt, Bruno}} \yr{2013}  \at{{Direct
  numerical simulations of local and global torque in Taylor--Couette flow up
  to Re= 30 000}}.  \jt{J. Fluid Mech.}  \bvol{718},  \pg{398--427}.

\bibitem[Chen {\em et~al.\/}(2022)Chen, Zhao \& Wan]{chen2022turbulent}
{\sc \au{Chen, Songtao}, \au{Zhao, Weiwen} \& \au{Wan, Decheng}} \yr{2022}
  \at{{Turbulent structures and characteristics of flows past a vertical
  surface-piercing finite circular cylinder}}.  \jt{Phys. Fluids}
  \bvol{34}~(1),  \pg{015115}.

\bibitem[Chouippe {\em et~al.\/}(2014)Chouippe, Climent, Legendre \&
  Gabillet]{chouippe2014numerical}
{\sc \au{Chouippe, Agathe}, \au{Climent, {\'E}ric}, \au{Legendre, Dominique} \&
  \au{Gabillet, C{\'e}line}} \yr{2014}  \at{{Numerical simulation of bubble
  dispersion in turbulent Taylor-Couette flow}}.  \jt{Phys. Fluids}
  \bvol{26}~(4).

\bibitem[Crialesi-Esposito {\em et~al.\/}(2022)Crialesi-Esposito, Rosti,
  Chibbaro \& Brandt]{crialesi-esposito_rosti_chibbaro_brandt_2022}
{\sc \au{Crialesi-Esposito, Marco}, \au{Rosti, Marco~Edoardo}, \au{Chibbaro,
  Sergio} \& \au{Brandt, Luca}} \yr{2022}  \at{{Modulation of homogeneous and
  isotropic turbulence in emulsions}}.  \jt{J. Fluid Mech.}  \bvol{940},
  \pg{A19}.

\bibitem[De~Vita {\em et~al.\/}(2019)De~Vita, Rosti, Caserta \&
  Brandt]{de2019effect}
{\sc \au{De~Vita, Francesco}, \au{Rosti, Marco~Edoardo}, \au{Caserta, Sergio}
  \& \au{Brandt, Luca}} \yr{2019}  \at{{On the effect of coalescence on the
  rheology of emulsions}}.  \jt{J. Fluid Mech.}  \bvol{880},  \pg{969--991}.

\bibitem[Dodd \& Ferrante(2016)]{dodd2016interaction}
{\sc \au{Dodd, Michael~S} \& \au{Ferrante, Antonino}} \yr{2016}  \at{{On the
  interaction of Taylor length scale size droplets and isotropic turbulence}}.
  \jt{J. Fluid Mech.}  \bvol{806},  \pg{356--412}.

\bibitem[Eckhardt {\em et~al.\/}(2007)Eckhardt, Grossmann \&
  Lohse]{eckhardt2007torque}
{\sc \au{Eckhardt, Bruno}, \au{Grossmann, Siegfried} \& \au{Lohse, Detlef}}
  \yr{2007}  \at{{Torque scaling in turbulent Taylor--Couette flow between
  independently rotating cylinders}}.  \jt{J. Fluid Mech.}  \bvol{581},
  \pg{221--250}.

\bibitem[Farsoiya {\em et~al.\/}(2023)Farsoiya, Liu, Daiss, Fox \&
  Deike]{farsoiya2023role}
{\sc \au{Farsoiya, Palas~Kumar}, \au{Liu, Zehua}, \au{Daiss, Andreas}, \au{Fox,
  Rodney~O} \& \au{Deike, Luc}} \yr{2023}  \at{{Role of viscosity in turbulent
  drop break-up}}.  \jt{J. Fluid Mech.}  \bvol{972},  \pg{A11}.

\bibitem[Favre(1969)]{favre1969statistical}
{\sc \au{Favre, A}} \yr{1969}  \at{{Statistical equations of turbulent gases}}.
   \jt{Problems of hydrodynamics and continuum mechanics}  \pg{pp. 231--266}.

\bibitem[Greenshields(2020)]{greenshields2020}
{\sc \au{Greenshields, Christopher}} \yr{2020} {\em OpenFOAM v8 User Guide\/}.
  \publ{London, UK: The OpenFOAM Foundation}.

\bibitem[Holzmann(2016)]{holzmann2016mathematics}
{\sc \au{Holzmann, Tobias}} \yr{2016}  \at{Mathematics, numerics, derivations
  and openfoam{\textregistered}}.  \jt{Loeben, Germany: Holzmann CFD} .

\bibitem[Hori {\em et~al.\/}(2023)Hori, Ng, Lohse \&
  Verzicco]{hori2023interfacial}
{\sc \au{Hori, Naoki}, \au{Ng, Chong~Shen}, \au{Lohse, Detlef} \& \au{Verzicco,
  Roberto}} \yr{2023}  \at{{Interfacial-dominated torque response in
  liquid--liquid Taylor--Couette flows}}.  \jt{J. Fluid Mech.}  \bvol{956},
  \pg{A15}.

\bibitem[Huisman {\em et~al.\/}(2013)Huisman, Scharnowski, Cierpka, K{\"a}hler,
  Lohse \& Sun]{huisman2013logarithmic}
{\sc \au{Huisman, Sander~G}, \au{Scharnowski, Sven}, \au{Cierpka, Christian},
  \au{K{\"a}hler, Christian~J}, \au{Lohse, Detlef} \& \au{Sun, Chao}} \yr{2013}
   \at{{Logarithmic boundary layers in strong Taylor-Couette turbulence}}.
  \jt{Phys. Rev. Lett.}  \bvol{110}~(26),  \pg{264501}.

\bibitem[Jim{\'e}nez(2012)]{jimenez2012cascades}
{\sc \au{Jim{\'e}nez, Javier}} \yr{2012}  \at{{Cascades in wall-bounded
  turbulence}}.  \jt{Annu. Rev. Fluid Mech.}  \bvol{44},  \pg{27--45}.

\bibitem[Kamp {\em et~al.\/}(2017)Kamp, Villwock \& Kraume]{kamp2017drop}
{\sc \au{Kamp, Johannes}, \au{Villwock, J{\"o}rn} \& \au{Kraume, Matthias}}
  \yr{2017}  \at{{Drop coalescence in technical liquid/liquid applications: A
  review on experimental techniques and modeling approaches}}.  \jt{Rev. Chem.
  Eng.}  \bvol{33}~(1),  \pg{1--47}.

\bibitem[Kilpatrick(2012)]{kilpatrick2012water}
{\sc \au{Kilpatrick, Peter~K}} \yr{2012}  \at{{Water-in-crude oil emulsion
  stabilization: review and unanswered questions}}.  \jt{Energy Fuels}
  \bvol{26}~(7),  \pg{4017--4026}.

\bibitem[Krieger \& Dougherty(1959)]{krieger1959mechanism}
{\sc \au{Krieger, Irvin~M} \& \au{Dougherty, Thomas~J}} \yr{1959}  \at{A
  mechanism for non-newtonian flow in suspensions of rigid spheres}.
  \jt{Trans. Soc. Rheol}  \bvol{3}~(1),  \pg{137--152}.

\bibitem[Lemenand {\em et~al.\/}(2017)Lemenand, Della~Valle, Dupont \&
  Peerhossaini]{lemenand2017turbulent}
{\sc \au{Lemenand, Thierry}, \au{Della~Valle, Dominique}, \au{Dupont, Pascal}
  \& \au{Peerhossaini, Hassan}} \yr{2017}  \at{{Turbulent spectrum model for
  drop-breakup mechanisms in an inhomogeneous turbulent flow}}.  \jt{Chem. Eng.
  Sci.}  \bvol{158},  \pg{41--49}.

\bibitem[Li {\em et~al.\/}(2006)Li, Sureshkumar \& Khomami]{li2006influence}
{\sc \au{Li, Chang-Feng}, \au{Sureshkumar, Radhakrishna} \& \au{Khomami,
  Bamin}} \yr{2006}  \at{{Influence of rheological parameters on polymer
  induced turbulent drag reduction}}.  \jt{J. Non-Newtonian Fluid Mech.}
  \bvol{140}~(1-3),  \pg{23--40}.

\bibitem[Mcclements(2007)]{mcclements2007critical}
{\sc \au{Mcclements, David~Julian}} \yr{2007}  \at{{Critical review of
  techniques and methodologies for characterization of emulsion stability}}.
  \jt{Crit. Rev. Food Sci. Nutr.}  \bvol{47}~(7),  \pg{611--649}.

\bibitem[Mukherjee {\em et~al.\/}(2019)Mukherjee, Safdari, Shardt,
  Kenjere{\v{s}} \& Van~den Akker]{mukherjee2019droplet}
{\sc \au{Mukherjee, Siddhartha}, \au{Safdari, Arman}, \au{Shardt, Orest},
  \au{Kenjere{\v{s}}, Sa{\v{s}}a} \& \au{Van~den Akker, Harry~EA}} \yr{2019}
  \at{{Droplet--turbulence interactions and quasi-equilibrium dynamics in
  turbulent emulsions}}.  \jt{J. Fluid Mech.}  \bvol{878},  \pg{221--276}.

\bibitem[Ni(2024)]{ni2023deformation}
{\sc \au{Ni, Rui}} \yr{2024}  \at{{Deformation and Breakup of Bubbles and Drops
  in Turbulence}}.  \jt{Annu. Rev. Fluid Mech.}  \bvol{56}~(1),  \pg{319--347}.

\bibitem[Ostilla {\em et~al.\/}(2013)Ostilla, Stevens, Grossmann, Verzicco \&
  Lohse]{ostilla2013optimal}
{\sc \au{Ostilla, Rodolfo}, \au{Stevens, Richard~JAM}, \au{Grossmann,
  Siegfried}, \au{Verzicco, Roberto} \& \au{Lohse, Detlef}} \yr{2013}
  \at{{Optimal Taylor--Couette flow: direct numerical simulations}}.  \jt{J.
  Fluid Mech.}  \bvol{719},  \pg{14--46}.

\bibitem[Ostilla-M{\'o}nico {\em et~al.\/}(2014)Ostilla-M{\'o}nico, Van
  Der~Poel, Verzicco, Grossmann \& Lohse]{ostilla2014boundary}
{\sc \au{Ostilla-M{\'o}nico, Rodolfo}, \au{Van Der~Poel, Erwin~P},
  \au{Verzicco, Roberto}, \au{Grossmann, Siegfried} \& \au{Lohse, Detlef}}
  \yr{2014}  \at{{Boundary layer dynamics at the transition between the
  classical and the ultimate regime of Taylor-Couette flow}}.  \jt{Phys.
  Fluids}  \bvol{26}~(1).

\bibitem[Ostilla-M{\'o}nico {\em et~al.\/}(2015)Ostilla-M{\'o}nico, Verzicco \&
  Lohse]{ostilla2015effects}
{\sc \au{Ostilla-M{\'o}nico, Rodolfo}, \au{Verzicco, Roberto} \& \au{Lohse,
  Detlef}} \yr{2015}  \at{{Effects of the computational domain size on direct
  numerical simulations of Taylor-Couette turbulence with stationary outer
  cylinder}}.  \jt{Phys. Fluids}  \bvol{27}~(2).

\bibitem[Perlekar {\em et~al.\/}(2014)Perlekar, Benzi, Clercx, Nelson \&
  Toschi]{perlekar2014spinodal}
{\sc \au{Perlekar, Prasad}, \au{Benzi, Roberto}, \au{Clercx, Herman~JH},
  \au{Nelson, David~R} \& \au{Toschi, Federico}} \yr{2014}  \at{Spinodal
  decomposition in homogeneous and isotropic turbulence}.  \jt{Phys. Rev.
  Lett.}  \bvol{112}~(1),  \pg{014502}.

\bibitem[Picano {\em et~al.\/}(2015)Picano, Breugem \&
  Brandt]{picano2015turbulent}
{\sc \au{Picano, Francesco}, \au{Breugem, Wim-Paul} \& \au{Brandt, Luca}}
  \yr{2015}  \at{{Turbulent channel flow of dense suspensions of neutrally
  buoyant spheres}}.  \jt{J. Fluid Mech.}  \bvol{764},  \pg{463--487}.

\bibitem[Rosti {\em et~al.\/}(2018)Rosti, Brandt \& Mitra]{rosti2018rheology}
{\sc \au{Rosti, Marco~E}, \au{Brandt, Luca} \& \au{Mitra, Dhrubaditya}}
  \yr{2018}  \at{{Rheology of suspensions of viscoelastic spheres:
  Deformability as an effective volume fraction}}.  \jt{Phys. Rev. Fluids}
  \bvol{3}~(1),  \pg{012301}.

\bibitem[Rosti {\em et~al.\/}(2019)Rosti, Ge, Jain, Dodd \&
  Brandt]{rosti2019droplets}
{\sc \au{Rosti, Marco~E}, \au{Ge, Zhouyang}, \au{Jain, Suhas~S}, \au{Dodd,
  Michael~S} \& \au{Brandt, Luca}} \yr{2019}  \at{{Droplets in homogeneous
  shear turbulence}}.  \jt{J. Fluid Mech.}  \bvol{876},  \pg{962--984}.

\bibitem[Rusche(2003)]{rusche2003computational}
{\sc \au{Rusche, Henrik}} \yr{2003}  \at{{Computational fluid dynamics of
  dispersed two-phase flows at high phase fractions}}. PhD thesis, Imperial
  College London (University of London).

\bibitem[Scheufler \& Roenby(2019)]{scheufler2019accurate}
{\sc \au{Scheufler, Henning} \& \au{Roenby, Johan}} \yr{2019}  \at{Accurate and
  efficient surface reconstruction from volume fraction data on general
  meshes}.  \jt{J. Comput. Phys.}  \bvol{383},  \pg{1--23}.

\bibitem[Soligo {\em et~al.\/}(2019)Soligo, Roccon \&
  Soldati]{soligo2019breakage}
{\sc \au{Soligo, Giovanni}, \au{Roccon, Alessio} \& \au{Soldati, Alfredo}}
  \yr{2019}  \at{Breakage, coalescence and size distribution of
  surfactant-laden droplets in turbulent flow}.  \jt{J. Fluid Mech.}
  \bvol{881},  \pg{244--282}.

\bibitem[Soligo {\em et~al.\/}(2021)Soligo, Roccon \&
  Soldati]{soligo2021turbulent}
{\sc \au{Soligo, Giovanni}, \au{Roccon, Alessio} \& \au{Soldati, Alfredo}}
  \yr{2021}  \at{Turbulent flows with drops and bubbles: what numerical
  simulations can tell us—freeman scholar lecture}.  \jt{J. Fluids Eng.}
  \bvol{143}~(8),  \pg{080801}.

\bibitem[Spandan {\em et~al.\/}(2016)Spandan, Ostilla-M{\'o}nico, Verzicco \&
  Lohse]{spandan2016drag}
{\sc \au{Spandan, Vamsi}, \au{Ostilla-M{\'o}nico, Rodolfo}, \au{Verzicco,
  Roberto} \& \au{Lohse, Detlef}} \yr{2016}  \at{{Drag reduction in numerical
  two-phase Taylor--Couette turbulence using an Euler--Lagrange approach}}.
  \jt{J. Fluid Mech.}  \bvol{798},  \pg{411--435}.

\bibitem[Spandan {\em et~al.\/}(2018)Spandan, Verzicco \&
  Lohse]{spandan2018physical}
{\sc \au{Spandan, Vamsi}, \au{Verzicco, Roberto} \& \au{Lohse, Detlef}}
  \yr{2018}  \at{{Physical mechanisms governing drag reduction in turbulent
  Taylor--Couette flow with finite-size deformable bubbles}}.  \jt{J. Fluid
  Mech.}  \bvol{849},  \pg{R3}.

\bibitem[Spernath \& Aserin(2006)]{spernath2006microemulsions}
{\sc \au{Spernath, Aviram} \& \au{Aserin, Abraham}} \yr{2006}
  \at{{Microemulsions as carriers for drugs and nutraceuticals}}.  \jt{Adv.
  Colloid Interface Sci.}  \bvol{128},  \pg{47--64}.

\bibitem[Su {\em et~al.\/}(2024)Su, Yi, Zhao, Wang, Xu, Wang \&
  Sun]{su2024numerical}
{\sc \au{Su, Jinghong}, \au{Yi, Lei}, \au{Zhao, Bidan}, \au{Wang, Cheng},
  \au{Xu, Fan}, \au{Wang, Junwu} \& \au{Sun, Chao}} \yr{2024}  \at{{Numerical
  study on the mechanism of drag modulation by dispersed drops in two-phase
  Taylor--Couette turbulence}}.  \jt{J. Fluid Mech.}  \bvol{984},  \pg{R3}.

\bibitem[Vela-Mart{\'\i}n \& Avila(2022)]{vela2022memoryless}
{\sc \au{Vela-Mart{\'\i}n, Alberto} \& \au{Avila, Marc}} \yr{2022}
  \at{Memoryless drop breakup in turbulence}.  \jt{Sci. Adv.}  \bvol{8}~(50),
  \pg{eabp9561}.

\bibitem[Wang {\em et~al.\/}(2023)Wang, Jiang \& Sun]{wang_jiang_sun_2023}
{\sc \au{Wang, Cheng}, \au{Jiang, Linfeng} \& \au{Sun, Chao}} \yr{2023}
  \at{{Numerical study on turbulence modulation of finite-size particles in
  plane-Couette flow}}.  \jt{J. Fluid Mech.}  \bvol{970},  \pg{A7}.

\bibitem[Wang {\em et~al.\/}(2022{\natexlab{{\em a\/}}})Wang, Yi, Jiang \&
  Sun]{wang2022finite}
{\sc \au{Wang, Cheng}, \au{Yi, Lei}, \au{Jiang, Linfeng} \& \au{Sun, Chao}}
  \yr{2022{\natexlab{{\em a\/}}}}  \at{{How do the finite-size particles modify
  the drag in Taylor--Couette turbulent flow}}.  \jt{J. Fluid Mech.}
  \bvol{937},  \pg{A15}.

\bibitem[Wang {\em et~al.\/}(2022{\natexlab{{\em b\/}}})Wang, Yi, Jiang \&
  Sun]{wang2022turbulence}
{\sc \au{Wang, Cheng}, \au{Yi, Lei}, \au{Jiang, Linfeng} \& \au{Sun, Chao}}
  \yr{2022{\natexlab{{\em b\/}}}}  \at{{Turbulence drag modulation by dispersed
  droplets in Taylor--Couette flow: the effects of the dispersed phase
  viscosity}}.  \jt{J. Fluid Mech.}  \bvol{952},  \pg{A39}.

\bibitem[Wang {\em et~al.\/}(2017)Wang, Abbas \& Climent]{wang2017modulation}
{\sc \au{Wang, Guiquan}, \au{Abbas, Micheline} \& \au{Climent, {\'E}ric}}
  \yr{2017}  \at{{Modulation of large-scale structures by neutrally buoyant and
  inertial finite-size particles in turbulent Couette flow}}.  \jt{Phys. Rev.
  Fluids}  \bvol{2}~(8),  \pg{084302}.

\bibitem[Wang {\em et~al.\/}(2007)Wang, Li, Zhang, Dong \& Eastoe]{wang2007oil}
{\sc \au{Wang, Lijuan}, \au{Li, Xuefeng}, \au{Zhang, Gaoyong}, \au{Dong,
  Jinfeng} \& \au{Eastoe, Julian}} \yr{2007}  \at{{Oil-in-water nanoemulsions
  for pesticide formulations}}.  \jt{J. Colloid Interface Sci.}
  \bvol{314}~(1),  \pg{230--235}.

\bibitem[Weller(2008)]{weller2008new}
{\sc \au{Weller, Henry~G}} \yr{2008}  \at{{A new approach to VOF-based
  interface capturing methods for incompressible and compressible flow}}.
  \jt{OpenCFD Ltd., Report TR/HGW}  \bvol{4},  \pg{35}.

\bibitem[Wong {\em et~al.\/}(2022)Wong, Baltzer, Livescu \&
  Lele]{wong2022analysis}
{\sc \au{Wong, Man~Long}, \au{Baltzer, Jon~R}, \au{Livescu, Daniel} \&
  \au{Lele, Sanjiva~K}} \yr{2022}  \at{{Analysis of second moments and their
  budgets for Richtmyer-Meshkov instability and variable-density turbulence
  induced by reshock}}.  \jt{Phys. Rev. Fluids}  \bvol{7}~(4),  \pg{044602}.

\bibitem[Xu {\em et~al.\/}(2023)Xu, Su, Lan, Zhao, He, Sun \&
  Wang]{xu2023direct}
{\sc \au{Xu, Fan}, \au{Su, Jinghong}, \au{Lan, Bin}, \au{Zhao, Peng}, \au{He,
  Yurong}, \au{Sun, Chao} \& \au{Wang, Junwu}} \yr{2023}  \at{Direct numerical
  simulation of {T}aylor--{C}ouette flow with vertical asymmetric rough walls}.
   \jt{J. Fluid Mech.}  \bvol{975},  \pg{A30}.

\bibitem[Xu {\em et~al.\/}(2022)Xu, Zhao, Sun, He \& Wang]{xu2022direct}
{\sc \au{Xu, Fan}, \au{Zhao, Peng}, \au{Sun, Chao}, \au{He, Yurong} \&
  \au{Wang, Junwu}} \yr{2022}  \at{Direct numerical simulation of
  {T}aylor-{C}ouette flow: Regime-dependent role of axial walls}.  \jt{Chem.
  Eng. Sci.}  \bvol{263},  \pg{118075}.

\bibitem[Yi {\em et~al.\/}(2021)Yi, Toschi \& Sun]{yi2021global}
{\sc \au{Yi, Lei}, \au{Toschi, Federico} \& \au{Sun, Chao}} \yr{2021}
  \at{{Global and local statistics in turbulent emulsions}}.  \jt{J. Fluid
  Mech.}  \bvol{912},  \pg{A13}.

\bibitem[Yi {\em et~al.\/}(2023)Yi, Wang, Huisman \& Sun]{yi2023recent}
{\sc \au{Yi, Lei}, \au{Wang, Cheng}, \au{Huisman, Sander~G} \& \au{Sun, Chao}}
  \yr{2023}  \at{{Recent developments of turbulent emulsions in Taylor--Couette
  flow}}.  \jt{Philos. Trans. R. Soc., A}  \bvol{381}~(2243),  \pg{20220129}.

\bibitem[Yi {\em et~al.\/}(2022)Yi, Wang, van Vuren, Lohse, Risso, Toschi \&
  Sun]{yi2022physical}
{\sc \au{Yi, Lei}, \au{Wang, Cheng}, \au{van Vuren, Thomas}, \au{Lohse,
  Detlef}, \au{Risso, Fr{\'e}d{\'e}ric}, \au{Toschi, Federico} \& \au{Sun,
  Chao}} \yr{2022}  \at{{Physical mechanisms for droplet size and effective
  viscosity asymmetries in turbulent emulsions}}.  \jt{J. Fluid Mech.}
  \bvol{951},  \pg{A39}.

\bibitem[Zhu {\em et~al.\/}(2016)Zhu, Ostilla-M{\'o}nico, Verzicco \&
  Lohse]{zhu2016direct}
{\sc \au{Zhu, Xiaojue}, \au{Ostilla-M{\'o}nico, Rodolfo}, \au{Verzicco,
  Roberto} \& \au{Lohse, Detlef}} \yr{2016}  \at{{Direct numerical simulation
  of Taylor--Couette flow with grooved walls: torque scaling and flow
  structure}}.  \jt{J. Fluid Mech.}  \bvol{794},  \pg{746--774}.

\end{thebibliography}

\end{document}